\magnification=\magstep1
\tolerance 500
\rightline{TAUP 2546-99}
\rightline{24 December, 2000}
\vskip  2 true cm
\centerline{\bf Lax-Phillips Scattering Theory}
\centerline{\bf of a}
\centerline{\bf Relativistic Quantum Field Theoretical Lee-Friedrichs Model}
\centerline{\bf and}
\centerline{\bf Lee-Oehme-Yang-Wu Phenomenology}
\smallskip
\centerline{Y. Strauss and L.P. Horwitz\footnote{*}{Also at Department of
Physics, Bar Ilan University, Ramat Gan 52900, Israel. e-mail:larry@
ccsg.tau.ac.il}}
\centerline{School of Physics}
\centerline{Raymond and Beverly Sackler Faculty of Exact Sciences}
\centerline{Tel Aviv University}
\centerline{Ramat Aviv 69978, Israel}
\bigskip
{\it Abstract:\/}
The one-channel Wigner-Weisskopf survival amplitude may be dominated by
exponential type decay in
pole approximation at times not too short or too long, but, in the
two channel case, for example, the pole residues are not orthogonal,
and the pole approximation evolution does not correspond to a
semigroup (experiments on
the decay of the neutral $K$-meson system support the
semigroup evolution postulated by Lee, Oehme and Yang, and Yang and
Wu to very high accuracy).  The scattering theory of Lax and Phillips,
 originally
developed for classical wave equations, has been recently extended to
the description of the evolution of resonant states in the framework
of quantum theory.  The resulting evolution law of the unstable system
is that of a semigroup, and the resonant state is a well-defined
function in the Lax-Phillips Hilbert space.  In this paper we apply
this theory to a relativistically covariant quantum field theoretical
form of the (soluble) Lee model.  We show that this theory provides a
rigorous underlying basis for the Lee-Oehme-Yang-Wu construction.  
\vfill
\eject
\noindent
{\bf 1. Introduction.}
\par The theory of Lax and Phillips$^1$ (1967), originally developed for the 
description of resonances in electromagnetic or acoustic scattering phenomena,
has been  used as a framework for the construction of a description of
 irreversible resonant phenomena in the
 quantum theory$^{2-5}$ (which we will refer to as the quantum Lax-Phillips 
theory). This leads to a time evolution of resonant states
which is of semigroup type, i.e., essentially exponential decay.  Semigroup
evolution is necessarily a property of irreversible processes$^6$. It 
appears experimentally that elementary particle decay, to a high 
degree of accuracy, follows a semigroup law, and hence such processes seem 
to be irreversible.    
\par The theory of Weisskopf and Wigner$^7$, which 
is based on the definition of the survival amplitude of the initial state
$\phi$ (associated with the unstable system) as the scalar 
product of that state
with the unitarily evolved state, 
$$  (\phi, e^{-iHt}\phi) \eqno(1.1)$$
cannot have exact exponential behavior$^8$. One can easily generalize this 
construction to the problem of more than one resonance$^{9,10}$. If $P$ is the 
projection operator into the  subspace of initial states
($N$-dimensional for $N$ resonances),
the reduced evolution operator is given by
$$Pe^{-iHt} P .\eqno(1.1')$$
\par Since the Laplace transform of this operator has a cut and not
just poles, this operator cannot be an element of an exact semigroup.$^8$
\par  Experiments on the decay of 
the neutral $K$-meson system$^{11}$ show clearly that the 
phenomenological description of Lee, Oehme and Yang$^{12}$, and Wu and 
Yang$^{13}$, by means of a $2\times 2$ effective Hamiltonian which corresponds
to an exact semigroup evolution of the unstable system, provides a very
accurate description of the data. Is can be proved that the Wigner-Weisskopf
theory cannot provide a semigroup evolution law$^8$ and, thus, the
effective $2\times 2$ Hamiltonian cannot emerge in the framework of this
theory. Furthermore, it has been shown, using estimates based on
the quantum mechanical Lee-Friedrichs model$^{14}$, that the experimental
results appear to rule out the application of the Wigner-Weisskopf theory to
the decay of the neutral $K$-meson system. While the exponential decay
law can be exhibited explicitly in terms of a Gel'fand
triple$^{15}$ (rigged Hilbert space), the 
representation of the resonant state in this framework is in a space
which does not coincide with the quantum mechanical Hilbert space, and
 does not have the properties of a Hilbert space, such as scalar products 
and the possibility of calculating physical properties associated
with expectation values.

\par The seminal work of Lax and Phillips$^1$ has provided us with the
 basic ideas necessary for the construction of a fundamental theoretical 
description, in the framework of the quantum theory$^{2-5}$,  of a
 resonant system which has exact semigroup evolution,
and represents the resonance as a {\it state in a Hilbert space}. In the 
following, we describe briefly the structure of this theory, a rather
straightforward generalization of standard quantum scattering theory,
and give
some physical interpretation for the states of the Lax-Phillips Hilbert space. 
\par The Lax-Phillips theory is defined in a Hilbert space $\overline H$
 of states which 
contains two distinguished subspaces, $ D_\pm$, called ``outgoing'' and 
``incoming''.  There is a unitary evolution law which we denote
 by $U(\tau)$, for which these subspaces are invariant in the following
sense:
$$\eqalign{ U(\tau)  D_+ &\subset  D_+ \qquad  \tau \geq 0 \cr
U(\tau) D_- &\subset  D_- \qquad \tau \leq 0 \cr} \eqno(1.2)$$
\par The translates of $ D_\pm$ under $U(\tau)$ are dense, i.e.,
$$ {\overline {{\bigcup_\tau}\, U(\tau)  D_\pm}} =  {\overline  H}
  \eqno(1.3)$$
and the asymptotic property
$$ {\bigcap_\tau }\,U(\tau)  D_\pm = \emptyset \eqno(1.4)$$
is assumed. It follows from these properties that 
$$ Z(\tau) = P_+ U(\tau) P_-, \eqno(1.5)$$
where $P_\pm$ are projections into the subspaces orthogonal to ${\cal D}_\pm$,
is a strongly contractive semigroup$^1$, i.e.,
$$ Z(\tau_1) Z(\tau_2) = Z(\tau_1 + \tau_2) \eqno(1.6)$$
for $\tau_1,\, \tau_2$ positive, and $\Vert Z(\tau)\Vert \to 0$ for $\tau \to
0$. It follows from $(1.2)$ that $Z(\tau)$ takes the subspace
$\cal K$, the orthogonal complement of $ D_\pm$ in $\overline H$
(associated with the resonances in the Lax-Phillips theory),
into itself$^1$, {\it i.e.},
$$ Z(\tau)= P_{\cal K} U(\tau) P_{\cal K}. \eqno(1.7)$$
The relation $(1.7)$ is of the same structure as $(1.1')$; there is, as we 
shall see in the following, an essential difference in the way that the
subspaces associated with resonances are defined. The argument that $(1.1')$
cannot form a semigroup is not valid$^3$ for $(1.7)$; the generator of
 $U(\tau)$ restricted to ${\cal K}$ is not self-adjoint. 
\par  A Hilbert 
space with the properties that there are distinguished subspaces 
 satisfying,
with a given law of evolution  $U(\tau)$, the properties $(1.2),\,(1.3),
\, (1.4)$ has a foliation$^{16}$ into a one-parameter (which we shall
 denote as $s$)
family of isomorphic Hilbert spaces, which are called {\it auxiliary}
Hilbert spaces, $\{ H_s\}$ for which
$$ {\overline  H} = {\int_\oplus}  H_s. \eqno(1.8)$$
Representing these spaces in terms of square-integrable functions, we define
the norm in the direct integral space 
as 
$$ \Vert f \Vert^2 = \int_{-\infty}^\infty ds \Vert f_s\Vert^2_H, \eqno(1.9)$$
where $f \in {\overline H} $ represents a vector in the
$L^2$ function space ${\overline H}=L^2(-\infty, \infty, H)$;  $f_s$ is 
an element of $H$, the $L^2$ 
function space (the {\it auxiliary space})
cooresponding to $ H_s$  for any $s$ [we shall not add in what follows a
subscript to the norm or scalar product symbols for scalar products of elements
of the auxiliary Hilbert space associated to a point $s$ on the foliation
axis since these spaces are all iso
morphic].
\par There are representations
for which the action of the full evolution group $U(\tau)$ on 
$L^2(-\infty, \infty;H)$ is translation by $\tau$ units. Given $D_\pm$
there is such a representation,
called the {\it incoming representation}$^1$, for which the set of all
functions in $D_-$ have support in $(-\infty, 0)$ and constitute the subspace
 $L^2(-\infty,0;H)$ of $L^2(-\infty, \infty;H)$; there is another
 representation,
called the {\it outgoing representation}, for which functions in $D_+$
have support in $(0,\infty)$ and 
constitute the subspace $L^2(0,\infty;H)$ of $L^2(-\infty, \infty;H)$.
The fact that $Z(\tau)$ in Eq. (1.7) is a semigroup is a consequence of
the definition of the subspaces $D_\pm$ in terms of support properties
on intervals along the foliation axis in the {\it outgoing} and {\it incoming}
translation representations respectively. The non-self-adjoint character of
the generator of the semigroup $Z(\tau)$ is a consequence of this
structure.$^3$ 
\par Lax and Phillips$^1$ show that there are unitary operators $W_\pm$, 
called wave operators, which map elements in ${\overline H}$
 to these representations.  They define an $S$-matrix, 
$$ S= W_+W_-^{-1}  \eqno(1.10)$$
which connects the incoming to the outgoing representations; it is
unitary, commutes with 
translations, and maps $L^2(-\infty,0;H)$ into itself. Since $S$
commutes with translations, it is diagonal in Fourier (spectral)
representation.  As pointed out by
Lax and Phillips$^1$, according to a special case of a theorem of
Four\`es and Segal$^{17}$, an operator with these properties can be
represented
as a multiplicative operator-valued function $S(\sigma)$ which
maps $H$ into itself, and satisfies the following conditions:
$$\eqalign{ (a)\   & S(\sigma)\  is\  the\  boundary\  
value\  of\  an\ \cr
&operator{\rm -}valued\  function\  S(z)\ analytic\ 
for\ {\rm Im}z >0. \cr
(b)\   &\Vert  S(z) \Vert  \leq 1\ for\ all\ z\ with\  {\rm Im}z >0. \cr
(c)\   & S(\sigma)\ is\ unitary\ for\ almost\ all\ real\ \sigma.\cr}$$ 
 An operator with
these properties is known as an inner function$^{18}$; such operators
arise in the study of shift invariant subspaces, the
essential mathematical content of the Lax-Phillips theory. 
 The singularities of
 this $S$-matrix, in what is called 
{\it spectral representation} (defined in terms of the Fourier
transform on the foliation variable $s$), correspond to the spectrum
of the generator of the semigroup characterizing the evolution 
of the unstable system.    
\par  In the framework of quantum theory, one may identify the Hilbert
space $ H$ with a space of physical states, and the variable
$\tau$ with the laboratory time (the semigroup
evolution is observed in the laboratory according to this time).
  The representation
of this space in terms of the foliated $L^2$ space ${\overline H}$
provides a natural probabilistic interpretation for the auxiliary spaces 
associated with each value of the foliation variable $s$, i.e., the 
quantity  $\Vert f_s \Vert^2 $ corresponds to the probability density
for the system to be found in the neighborhood of $s$. For example,
consider an operator $A$ defined on ${\overline H}$ which acts pointwise, i.e.,
contains no shift along the foliation. Such an operator can be
represented as a direct integral
$$ A = \int_\oplus A_s.  \eqno(1.11)$$
It produces a map
of the auxiliary space $H$ into $H$ for each value of $s$, and thus,
if it is self-adjoint, $A_s$  may act as an observable in a quantum theory
associated to the point $s$$^{4}$; The expectation value of $A_s$ in a state
in this Hilbert space defined by the vector $\psi_s$, the component of
$\psi \in {\overline H}$ in the auxiliary space at $s$, is 
$$ \langle A_s \rangle_s = {(\psi_s, A_s \psi_s) \over \Vert \psi_s
\Vert^2}
 \eqno(1.12) $$.
Taking into account the {\it a priori} probability density $\Vert
\psi_s\Vert^2$ that the system is found at this point on the foliation
axis, we see that the expectation value of $A$ in ${\overline H}$ is
$$ \langle A \rangle =  \int ds \langle A_s \rangle_s \Vert \psi_s\Vert^2
= \int ds (\psi_s, A_s \psi_s ), \eqno(1.13)$$   
the direct integral representation of $(\psi, A \psi)$.
\par As we have remarked above, in the translation representations for
 $U(\tau)$ the foliation variable $s$ is shifted (this shift, for
 sufficiently large $\vert \tau \vert$,  induces the transition of
 the state into the subspaces $ D_\pm$). It follows that $s$
 may be identified as an intrinsic time associated with the evolution
 of the state; since it is a variable of the measure space of the
 Hilbert space ${\overline H}$, this quantity itself has the
 meaning of a quantum variable.   
\par We are presented here with the notion of a virtual history. To
understand this idea, suppose that at a given time $\tau_0$, the
function which represents the state has some distribution $\Vert
\psi_s^{\tau_0}\Vert^2$. This distribution provides an {\it a priori}
probability that the system would be found at time $s$ (greater or less than
$\tau_0$), if the experiment were to be performed at time $s$ corresponding 
 to $\tau = s$ on the laboratory clock.  The state of the system
therefore contains information on the structure of the {\it history} of
 the system as it is inferred at $\tau_0$.   
\par We shall assume the existence of a unitary evolution on the
Hilbert space $\overline{H}$, and that for
$$ U(\tau) = e^{-iK\tau}, \eqno(1.14)$$ 
the generator $K$ can be decomposed as 
$$ K= K_0 + V \eqno(1.15)$$
 in terms of an unperturbed
operator $K_0$ with spectrum $(-\infty, \infty)$ and a perturbation
$V$, under which this spectrum is stable.  
We shall, furthermore, assume that wave operators exist, defined on
some dense set, as 
$$ \Omega_\pm = \lim_{\tau \rightarrow \pm \infty} e^{iK\tau}e^{-iK_0
\tau}. \eqno(1.16)$$
In the soluble model that we shall treat as an example in this paper,
 the existence of the wave operators is assured.
\par  With the help of the
 wave operators, we can define translation representations for $U(\tau)$. 
 The translation representation for $K_0$ is defined by the property
$$ {{}_0\langle} s, \alpha \vert e^{-iK_0\tau} f)= {{}_0\langle} s-\tau, \alpha
\vert f), \eqno(1.17)$$
where $\alpha$ corresponds to a label for the basis of the auxiliary space.  
Noting that
$$  K \Omega_\pm = \Omega_\pm K_0 \eqno(1.18) $$
we see that
$$ {{}_{out \atop in}\langle} s, \alpha \vert e^{-iK\tau} f) =
 {{}_{out \atop in} \langle}
s-\tau, \alpha \vert f),  \eqno(1.19)$$
where
$$ {{}_{out \atop in}\langle}s, \alpha \vert f) = {{}_0 \langle}s,
  \alpha \vert 
\Omega_\pm^\dagger f) \eqno(1.20)$$
\par It will be convenient to work in terms of the Fourier transform of
 the {\it in} and {\it out} 
translation representations; we shall call these  the {\it in} and
 {\it out} {\it spectral}
representations, {\it i.e.},
$$ {{}_{out \atop in}\langle} \sigma, \alpha \vert f) = \int_{-\infty}^\infty
e^{-i\sigma s} {{}_{out \atop in}\langle} s, \alpha \vert f). \eqno(1.21)$$
In these representations, (1.20) is
$$ {{}_{out \atop in}\langle}\sigma, \alpha \vert f) = {{}_0 \langle}\sigma,
   \alpha \vert \Omega_\pm^\dagger f)
    \eqno(1.22)$$
and $(1.19)$ becomes
$$ {{}_{out \atop in}\langle} \sigma, \alpha \vert e^{-iK\tau}\vert f) =
  e^{-i\sigma \tau}{{}_{out \atop in}\langle} \sigma, \alpha \vert f).
\eqno(1.23)$$
Eq. $(1.17)$ becomes, under Fourier transform
$$ {{}_0\langle} \sigma, \alpha \vert e^{-iK_0\tau} f)= e^{-i\sigma \tau}
{{}_0\langle} \sigma, \alpha \vert f). \eqno(1.24)$$
For $f$ in the domain of $K_0$, $(1.23)$ implies that
$$ {{}_0\langle} \sigma, \alpha \vert K_0 f)= \sigma
{{}_0\langle} \sigma, \alpha \vert f). \eqno(1.25)$$
\par With the solution of $(1.25)$, and the wave operators,
the {\it in} and {\it out} spectral 
representations of a vector $f$ can be constructed from $(1.24)$.
\par We are now in a position to construct the subspaces
 $D_\pm$, which are not given {\it a priori} (as they are in the
classical theory$^1$) in the Lax-Phillips quantum
theory.  We shall define $ D_+$ as the
set of functions with  support in $(0,\infty)$ in the {\it outgoing}
translation representation. Similarly, we shall define $ D_-$ as the
set of functions with  support in $(-\infty,0)$ in the {\it incoming}
translation representation.
representation. The corresponding elements of ${\overline H}$
constitute the subspaces $ D_\pm$.   By construction,
$ D_\pm$ have the required invariance properties under the
action of $U(\tau)$.
\par  The {\it outgoing spectral representation}
of a vector $g \in {\cal H}$ is
$$\eqalign{  {{}_{out}\langle} \sigma \alpha \vert g ) =
{{}_0\langle}\sigma\alpha \vert \Omega_+^{-1} g)&= 
\int\,d\sigma' \sum_{\alpha'} {{}_0 \langle} \sigma \alpha \vert {\bf
S}\vert \sigma' \alpha' \rangle_0\,\, {{}_0 \langle}\sigma' \alpha' \vert
\Omega_-^{-1} g)\cr & = \int\,d\sigma' \sum_{\alpha'} {{}_0 \langle} 
\sigma \alpha \vert {\bf
S}\vert \sigma' \alpha' \rangle_0 \,\,{{}_{in} \langle}\sigma' \alpha' 
\vert g),\cr} \eqno (1.26)$$
where we call 
$${\bf S} = \Omega_+^{-1} \Omega_-. \eqno(1.27)$$
the quantum Lax-Phillips $S$-operator. We see that the kernel ${{}_0 \langle} 
\sigma \alpha \vert {\bf
S}\vert \sigma' \alpha' \rangle_0$ maps the incoming to outgoing
spectral representations.  Since $\bf S$ commutes with $K_0$, it
follows that 
$$ {{}_0 \langle} 
\sigma \alpha \vert {\bf
S}\vert \sigma' \alpha' \rangle_0 = \delta(\sigma - \sigma') S^{\alpha
\alpha'}. (\sigma) \eqno(1.28)$$
It follows from $(1.16)$ and $(1.22)$, in the standard way$^{19}$, that
$$ {{}_0\langle} \sigma \alpha \vert {\bf S} \vert \sigma' \alpha' 
{\rangle_0} =  \lim_{\epsilon \rightarrow 0}
\delta(\sigma-\sigma')\{\delta^{\alpha \alpha'} - 
2\pi i {{}_0\langle} \sigma \alpha \vert {\bf T}(\sigma + i\epsilon)
 \vert \sigma' \alpha' {\rangle_0} \}, \eqno(1.29)$$
where
$$ {\bf T}(z) = V + VG(z)V = V + VG_0(z) {\bf T}(z). \eqno(1.30)$$
We remark that, by this construction,
$S^{\alpha \alpha'}(\sigma)$ is {\it analytic in the upper half plane} in 
$\sigma$.
The Lax-Phillips $S$-matrix$^1$ is given by the inverse Fourier transform,
$$ S = \bigl\{{{}_0 \langle} 
s \alpha \vert {\bf
S}\vert s' \alpha' \rangle_0\bigr\};  \eqno(1.31)$$
this operator clearly commutes with translations. 
\par From $(1.29)$ it follows that the inner function property $(a)$ 
 of $S(\sigma)$ above is true. Property 
$(c)$, unitarity for real $\sigma$, is equivalent to asymptotic
completeness, a property which is stronger than the existence of wave
operators. For the relativistic Lee model, which  we shall treat in
this paper, this condition is satisfied.   In the model that we shall 
study here, we shall see that there is a wide class of potentials $V$ for
which the operator $S(\sigma)$ satisfies the property $(b)$.
\par In the next section, we review briefly the structure of the 
relativistic Lee model$^{19}$, and construct explicitly the
Lax-Phillips spectral representations and $S$-matrix. Introducing
auxiliary space variables, we then characterize the properties of the
finite rank Lee model potential which assure that the $S$-matrix is an
inner function, {\it i.e.}, that property $(b)$ listed above  is satisfied.
\bigskip
\par{\bf 2. The multi-channel relativistic Lee-Friedrichs model}
\smallskip 
\par The multi-channel relativistic Lee-Friedrichs model is defined in terms
of bosonic quantum fields on spacetime. These fields, which emerge
from the second quantization of the Stueckelberg covariant quantum
theory$^{20}$, evolve with an invariant
evolution parameter$^{5}$ $\tau$ (which we identify here with the
evolution parameter of the Lax-Phillips theory discussed above); at equal
$\tau$, they satisfy the commutation relations (with $\psi_i^\dagger$ as the
canonical conjugate field to $\psi_i$; the fields $\psi_i$, which satisfy first
order evolution equations as for nonrelativistic Schrodinger fields, are
just annihilation operators)

$$ [\psi_{i\tau}(x), \psi^\dagger_{j \tau}(x')]=\delta^4(x-x')\delta_{ij}.
 \eqno(2.1)$$

\par Transforming to momentum space, in which we have

$$ \psi_{i\tau}(p)={1\over (2\pi)^2}\int d^4x\, e^{-ip_\mu x^\mu}
   \psi_{i\tau}(x), \eqno(2.2)$$
relation (2.1) becomes

$$ [\psi_{i\tau}(p),\psi_{j\tau}^\dagger(p')]=\delta^4(p-p')\delta_{ij}. \eqno(2.3)$$
The manifestly covariant spacetime structure of these fields is admissible
when $E,{\bf p}$ are not {\it a priori} constrained by a sharp
mass-shell relation. In the mass-shell limit (for which the variation in
$m^2$ defined by $E^2-{\bf p}^2$ is small), multiplying both sides of (2.3)
by $\Delta E=\Delta m^2/2E$, one obtains the usual commutation relations
for on shell fields,

$$ [\tilde \psi_{i\tau} ({\bf p}),\tilde \psi_{j\tau}^\dagger ({\bf p})]=2E
   \delta^3({\bf p}-{\bf p'})\delta_{ij}, \eqno(2.4)$$
where $\tilde\psi_{i\tau}({\bf p})=\sqrt{\Delta m^2}\psi_{i\tau}(\bf p)$. 
In this limit, $t$ and $\tau$ coincide.  The generator of
evolution
$$ K=K_0+V \eqno(2.5)$$
for which the Heisenberg picture fields are

$$ \psi_{i\tau}(p)=e^{iK\tau}\psi_{i0}(p)e^{-iK\tau} \eqno(2.6)$$
is given, in this model, as (we write $p^2=p_\mu p^\mu, k^2=k_\mu k^\mu$
in the following)\footnote{*}{We remark that Antoniou, {\it et al}
$^{21}$, have constructed a
relativistic Lee model of a somewhat different type; their field equation
is second order in derivative with respect to the variable conjugate to the
mass.}

$$ \eqalign{K_0=\sum_{i=1,2} \Big\{ \int d^4p\, {p^2\over 2M_{V_i}}
   b_i^\dagger(p)b_i(p)
   &+\int d^4p\, {p^2\over 2M_{N_i}} a^\dagger_{N_i}(p)a_{N_i}(p)\Big\}\cr
   &+\sum_{i=1,2}\int d^4p\,
   {p^2\over 2M_{\theta_i}}a_{\theta_i}^\dagger(p)a_{\theta_i}(p)\cr}
   \eqno(2.7)$$
and

$$ V=\sum_{i,j=1,2} \int d^4p\ \int d^4k\
   \Bigl( f_{ij}(k)b_i^\dagger(p)a_{N_j}(p-k)
   a_{\theta_j}(k)+f_{ij}^*(k)b_i(p)a_{N_j}^\dagger(p-k)a_{\theta_j}^\dagger
   (k)\Bigr) \eqno(2.8) $$

This model describes the process $V_i \rightarrow N_j+\theta_j$. We
assume that the fields associated with different particles commute. The fields
$b_i(p), a_{N_i}(p)$ and $a_{\theta_i}$ are annihilation operators for the
$V_i,N_i$ and $\theta_i$ particles, respectively. We take $M_{V_i},M_{N_i}$
and $M_{\theta_i}$ to be the mass parameters for the fields$^{19,22}$. We restrict our development to the two channel case in the
following.  The generalization to any number of channels is straightforward.
\par The following operators are conserved

$$ \eqalign{Q_1 &=\sum_{i=1,2}\int d^4p\ (b_i^\dagger(p)b_i(p)
   +a_{N_i}^\dagger(p)a_{N_i}(p)) \cr
   Q_2&=\int d^4p\ (a_{N_1}^\dagger(p)
   a_{N_1}(p) - a_{\theta_1}^\dagger(p)a_{\theta_1}(p) ) \cr
   Q_3&=\int d^4p\ (a_{N_2}^\dagger(p)a_{N_2}(p) - a_{\theta_2}^\dagger(p)
   a_{\theta_2}(p) )\cr}. \eqno(2.9) $$
This fact enables us to decompose the Fock space to sectors. We 
study the sector with $Q_1=1,Q_2=0,Q_3=0$. This is identified as a sector
containing either one $V_i$ particle {\it or} one $N_j$ together with one
$\theta_j$ particle. It follows from the commutativity of the fields
that the states $\vert V_1 \rangle,\, \vert V_2 \rangle$, as well as
$\vert N_1 \theta_1 \rangle,\,  \vert N_2, \theta_2 \rangle$, which
exist in this sector, are orthogonal. In this sector the generator of
evolution $K$ can be rewritten in the form

$$ K=\int d^4p\, K^p=\int d^4p\, (K_0^p+V^p)$$
where

$$ K_0^p=\sum_{i=1,2}\Bigl\{ {p^2\over 2M_{V_i}}b_i^\dagger(p)b_i(p)
   +\int d^4k\left( {(p-k)^2\over 2M_{N_i}}+{k^2\over 2M_{\theta_i}}
   \right)a_{N_i}^\dagger(p-k)a_{\theta_i}^\dagger(k)a_{\theta_i}(k)
   a_{N_i}(p-k)\Bigr\}$$
and

$$ V^p=\sum_{i,j=1,2}^n \int d^4k\
   \big( f_{ij}(k)b_i^\dagger(p)a_{N_j}(p-k)
   a_{\theta_j}(k)+f_{ij}^*(k)b_i(p)a_{N_j}^\dagger(p-k)a_{\theta_j}^\dagger
   (k)\big) $$
In this form it is clear that both $K$ and $K_0$ have a direct integral
structure. This implies a similar structure for the wave operator
$\Omega_\pm$ and the possibility of defining restricted wave operators
$\Omega_\pm^p$ for each value of p. We see from the expression for $K_0^p$
that $\vert V_i(p)\rangle=b_i^\dagger(p)\vert 0\rangle$ can be
regarded as a set of discrete eigenstates of $K_0^p$ (for each $p$)
which span a subspace
which is, therefore, annihilated by the restricted wave operators
$\Omega_\pm^p$. This implies immediately that $\Omega_\pm\vert V(p)\rangle=0$
for every p (an explicit demonstration of this fact is given in appendix A).
\par In order to construct the Lax-Phillips incoming and outgoing spectral
representations for the model presented here it is necessary, according to
the discussion following equation (1.25), to obtain appropriate expressions
for the wave operators $\Omega^\dagger_\pm$ and the spectral representation
for the generator $K_0$ of free evolution, i.e, the solution of equation
(1.25).
\par We begin our discussion with a derivation of the appropriate expressions,
for the model considered here, of the wave operators $\Omega_\pm$. We first
calculate the following matrix elements of $\Omega_+$

$$ \langle V_m(q) \vert \Omega_+ \vert N_n(p),\theta_n(k)\rangle \qquad
   \langle N_m(p'),\theta_m(k')\vert \Omega_+\vert N_n(p),\theta_n(k)
   \rangle $$
Equation (1.16) can be rewritten, following the standard procedure$^{23}$,
in the integral form

$$ \Omega_+ = 1+i\lim_{\epsilon\to 0}\int _0^{+\infty} U^\dagger(\tau)
   VU_0(\tau)e^{-\epsilon\tau}d\tau \eqno(2.10) $$
where $U(\tau)=e^{-iK\tau},\ U_0(\tau)=e^{-iK_0\tau}$. Using (2.7), we have

$$ \Omega_+\vert N_n(p_1),\theta_n(p_2)\rangle
   =\vert N_n(p_1),\theta_n(p_2)
   \rangle+i\lim_{\epsilon\to 0}\int_0^{+\infty}d\tau\,
   e^{-\epsilon\tau}U^\dagger(\tau)VU_0(\tau)a^\dagger_{N_n}(p)
   a^\dagger_{\theta_n}(k)\vert 0\rangle$$
$$ =\vert N_n(p_1),\theta_n(p_2) \rangle
   +i \lim_{\epsilon\to 0}
   \int_0^{+\infty} d\tau e^{-i(\omega_{N_n}(p_1)+\omega_{\theta_n}(p_2)
   -i\epsilon)\tau} U^\dagger(\tau)V a^\dagger_{N_n}(p)
   a^\dagger_{\theta_n}(k)\vert 0\rangle \eqno(2.11)$$
where $\omega_{N_n}(p)=p^2/2M_{N_n},\ \omega_{\theta_n}(p)=p^2/2M_{\theta_n}$.
Using (2.8) we find 

$$ V a_{N_n}^\dagger(p_1)a_{\theta_n}^\dagger(p_2)\vert 0 \rangle
   =\sum_{k=1,2} f_{kn}(p_2)b_k^\dagger(p_1+p_2)\vert 0 \rangle \eqno(2.12)$$
Inserting (2.12) into (2.11) and changing the integration variable from $\tau$
to $-\tau$ it follows that

$$ \Omega_+\vert N_n(p_1),\theta_n(p_2)\rangle=$$
$$ =\vert N_n(p_1),\theta_n(p_2) \rangle
   -i \lim_{\epsilon\to 0}
   \sum_{k=1,2}\int_0^{-\infty} d\tau e^{i(\omega_{N_n}(p_1)+\omega_{\theta_n}
   (p_2)-i\epsilon)\tau} U(\tau)
   f_{kn}(p_2)b_k^\dagger(p_1+p_2)\vert 0 \rangle \eqno(2.13)$$
In order to continue with the evaluation of the integral in $(2.13)$ we find
the time evolution of some arbitray state $\chi$ under the action of $U(\tau)$

$$ \psi(\tau)=U(\tau)\chi =e^{-iK\tau}\chi \eqno(2.14)$$
In the sector of the Fock space that we are considering, the state
$\psi(\tau)$ at any time $\tau$ can be expanded as

$$ \psi(\tau) = \sum_{i=1,2}\int d^4q\ A_i(q,\tau)b_i^\dagger(q)
   \vert 0 \rangle + \sum_{i=1,2}\int d^4p\ \int d^4k\ B_i(p,k,\tau)
   a_{N_i}^\dagger(p) a_{\theta_i}^\dagger(k)\vert 0 \rangle \eqno(2.15) $$
In particular, we see that the initial conditions for the evolution in
$(2.13)$, where the state $\chi$ is taken to be $\psi_0=\sum_k f_{kn}(p_2)
b^\dagger_k(p_1+p_2)\vert 0 \rangle$, are

$$ A_i(q,0)=f_{in}(p_2)\delta^4(q-p_1-p_2) \qquad  B_i(p,k,0)=0 \eqno(2.16)$$

The equations of evolution for the coefficients $A(q,\tau)$ and $B(p,k,\tau)$
are then obtained from (2.14) and (2.15), {\it i.e.},

$$ \eqalign{i{\partial B_i(p-k,k,\tau)\over{\partial \tau}}&=B_i(p-k,k,\tau)
   \left({(p-k)^2\over{2M_{N_i}}}+{k^2\over{2M_{\theta_i}}}\right)
   +\sum_{j=1,2}f_{ji}^*(k)A_j(p,\tau)\cr
   i{\partial A_i(p,\tau)\over{\partial \tau}}&={p^2\over{2M_{V_i}}}
   A_i(p,\tau)+\sum_{j=1,2}\int d^4k f_{ij}(k)B_j(p-k,k,\tau)\cr}
   \eqno(2.17)$$
These equations can be solved algebraically$^{10,19}$ by performing Laplace
transforms and defining

$$ \eqalign{\tilde B_i(p,k,z)&=\int_{-\infty}^0 d\tau e^{iz\tau}
   B_i(p,k,\tau)\qquad  {\rm Im} z<0 \cr
   \tilde A_i(p,z)&=\int_{-\infty}^0 d\tau e^{iz\tau} A_i(p,\tau)
   \qquad {\rm Im} z<0 \cr}.\eqno(2.18)$$
Equations (2.17) are transformed into

$$ \eqalign{&\tilde B_i(p-k,k,z)\left(z-{(p-k)^2\over {2M_{N_i}}}-{k^2\over
   {2M_{\theta_i}}}\right)=iB_i(p-k,k,0)+\sum_{j=1,2}f_{ji}^*(k)
   \tilde A_j(p,z) \cr 
   &\tilde A_i(p,z)(z-{p^2\over {2M_{V_i}}})=iA_i(p,0)+\sum_{j=1,2}\int
   d^4k f_{ij}(k)\tilde B_j(p-k,k,z)\cr} \eqno(2.19)$$
Using the initial conditions (2.16) we obtain the following expressions for
the Laplace transformed coefficients

$$ \eqalign{\tilde A_k(p,z)&=i\sum_{i=1}^n W_{ki}^{-1}(z,p)A_i(p,0) \cr
   \tilde B_j(p-k,k,z)&=i\left({z-(p-k)^2\over{2M_{N_j}}}-{k^2\over
   {2M_{\theta_j}}}\right)^{-1}\left[\sum_{k,i=1,2}^nf_{kj}^*(k)W_{ki}^{-1}(z,p)
   A_i(p,0)\right]\cr} \eqno(2.20)$$
where

$$ W_{ik}(z,p)=\delta_{ik}(z-{p^2\over{2M_{V_i}}})-\sum_{j=1,2}^n\int d^4k
   {f_{ij}(k)f_{kj}^*(k)\over z-{(p-k)^2\over{2M_{N_j}}}-{k^2\over
   {2M_{\theta_j}}} } \eqno(2.21)$$
The Laplace transform of $\psi(\tau)$ is then

$$ \psi(z)= i\sum_{i,k=1,2}\int d^4q
   W^{-1}_{ik}(z,q)A_k(q,0)b^\dagger_i
   (q)\vert 0 \rangle$$
$$ +i\sum_{i,j,k=1,2}\int d^4p \int d^4k
   {f^*_{ki}(k)W^{-1}_{kj}(z,p+k)A_j(p+k,0)\over
   {\left( z-{{ p^2}\over 2M_{N_i}}
   -{{k^2}\over 2M_{\theta_i}} \right)}}
   a^\dagger_{N_i}(p)a^\dagger_{\theta_i}(k)
   \vert 0 \rangle \eqno(2.22) $$
From (2.13),(2.22) and the initial conditions Eq.(2.16) we get

$$ \Omega_+\vert N_n(p_1), \theta_n(p_2) \rangle=\vert N_n(p_1),
   \theta_n(p_2) \rangle
   +i\sum_{i,k=1,2} W^{-1}_{ik}(\omega_n-i\epsilon,p_1+p_2)f_{kn}(p_2)
   b^\dagger_i(p_1+p_2)\vert 0 \rangle$$
$$ +i\sum_{i,j,k=1,2}\int d^4k
   {{f^*_{ki}(k)W^{-1}_{kj}(\omega_n-i\epsilon,
   p_1+p_2)f_{jn}(p_2)}\over
   \omega_n-i\epsilon-{{(p_1+p_2-k)^2}\over
   2M_{N_i}}-{{k^2}\over 2M_{\theta_i}} }
   a^\dagger_{N_i}(p_1+p_2-k)a^\dagger_{\theta_i}(k)
   \vert 0 \rangle \eqno(2.23) $$
where we denote $\omega_n \equiv \omega_{N_n}(p_1)+\omega_{\theta_n}(p_2)$.
We can now evaluate the desired matrix elements of the wave operator
$\Omega_+$. From (2.23) one fimds

$$ \langle V_m(p) \vert \Omega_+\vert N_n(p_1), \theta_n(p_2) \rangle=
   \sum_{k=1,2} W^{-1}_{mk}(\omega_n-i\epsilon,p_1+p_2)f_{kn}(p_2)
   \delta^4(p-p_1-p_2) \eqno(2.24)$$

and
$$ \langle N_m(\tilde p_1),\theta_m(\tilde p_2)\vert \Omega_+\vert N_n(p_1),
   \theta_n(p_2) \rangle=\delta_{mn}\delta^4(\tilde p_1-p_1)
   \delta^4(\tilde p_2-p_2)$$
$$ +\sum_{k,j=1,2}
   {{f^*_{km}(\tilde p_2)W^{-1}_{kj}(\omega_n
   -i\epsilon,p_1+p_2)f_{jn}(p_2)}\over
   \omega_n-i\epsilon-{{\tilde p_1^2}\over
   2M_{N_m}}-{{\tilde p_2^2}\over 2M_{\theta_m}}}
   \delta^4(p_1+p_2-\tilde p_1-\tilde p_2)
   \eqno(2.25) $$
In a similar fashion one can find the corresponding matrix elements of the
wave operator $\Omega_-$:

$$ \langle V_m(p) \vert \Omega_-\vert N_n(p_1), \theta_n(p_2) \rangle=
   \sum_{k=1,2} W^{-1}_{mk}(\omega_n+i\epsilon,p_1+p_2)f_{kn}(p_2)
   \delta^4(p-p_1-p_2) \eqno(2.26)$$
and

$$ \langle N_m(\tilde p_1),\theta_m(\tilde p_2)\vert \Omega_-\vert N_n(p_1),
   \theta_n(p_2) \rangle=\delta_{mn}\delta^4(\tilde p_1-p_1)
   \delta^4(\tilde p_2-p_2)$$
$$ +\sum_{k,j=1,2}
   {{f^*_{km}(\tilde p_2)W^{-1}_{kj}(\omega_n
   +i\epsilon,p_1+p_2)f_{jn}(p_2)}\over
   \omega_n+i\epsilon-{{\tilde p_1^2}\over
   2M_{N_m}}-{{\tilde p_2^2}\over 2M_{\theta_m}}}
   \delta^4(p_1+p_2-\tilde p_1-\tilde p_2)
   \eqno(2.27) $$
\par According to eq. (1.22), (1.25) the complete transformation to the
{\it incoming} or {\it outgoing} representations requires us to solve for the
(improper) eigenvectors with spectrum $\{\sigma\}$ on $(-\infty,+\infty)$
of $K_0$. The complete set of these states is decomposed into two subsets
corresponding to the quantum numbers for states containing $N$ and $\theta$
particles and states containing a $V$ particle. These quantum numbers are
denoted $\sigma,\alpha$ (for the $N+\theta$ states)and $\sigma,\beta$
($V$ states) respectively. For the projections into these two subspaces,
we have

$$ \eqalign{\vert \sigma,\alpha\rangle_0&=\sum_{n=1,2}\int d^4p \int d^4k\,
   \vert N_l(p),\theta_l(k)\rangle \langle N_l(p),\theta_l(k) \vert
   \sigma,\alpha\rangle_0 \cr
   \vert \sigma,\beta\rangle_0&=\sum_{m=1}\int d^4p\, \vert V_m(p)\rangle
   \langle V_m(p)\vert \sigma,\beta \rangle_0 \cr} \eqno(2.28)$$
It is convenient to define

$$ \eqalign{O^{\sigma,\alpha}_{n,p,k}&\equiv \langle N_n(p),\Theta_n(k)\vert
   \sigma,\alpha \rangle_0\cr
   O^{\sigma,\beta}_{m,p}&\equiv \langle V_m(p)\vert\sigma,\beta \rangle_0
   \cr} \eqno (2.29)$$
With the help of these definitions we can rewrite (2.28) as

$$ \eqalign{\vert \sigma,\alpha\rangle_0&=\sum_{n=1,2}\int d^4p \int d^4k\ 
   O^{\sigma,\alpha}_{n,p,k}\vert N_l(p),\theta_l(k)\rangle \cr
   \vert \sigma,\beta\rangle_0&=\sum_{m=1,2}\int d^4p\ 
   O^{\sigma,\beta}_{m,p} \vert V_m(p)\rangle
    \cr} \eqno(2.30)$$
It follows from equations (1.25) and (2.30) that

$$ \eqalign{K_0\vert \sigma,\alpha\rangle_0&=\sum_{n=1,2}\int d^4p\int d^4k\ 
   (\omega_{N_n}(p)+\omega_{\theta_n}(k))O^{\sigma,\alpha}_{n,p,k}\vert N_n(p)
   ,\Theta_n(k)\rangle=\sigma\vert \sigma,\alpha\rangle_0 \cr
   K_0\vert \sigma,\beta\rangle_0&=\sum_{m=1,2}\int d^4p\ 
   \omega_{V_m}(p)O^{\sigma,\beta}_{m,p}\vert V_m(p)\rangle
   =\sigma\vert \sigma,\beta\rangle_0 \cr} \eqno(2.31) $$
From the orthogonality of the final state channels, it follows that we must
have

$$ \eqalign{O^{\sigma,\alpha}_{n,p,k}&=\delta(\sigma-\omega_{N_n}(p)
   -\omega_{\theta_n}(k))\tilde O^{\sigma,\alpha}_{n,p,k}\cr
   O^{\sigma,\beta}_{m,p}&=\delta(\sigma-\omega_{V_m}(p))
   \tilde O^{\sigma,\beta}_{m,p}\cr} \eqno(2.32) $$
to satisfy the kinematic conditions imposed by eq. (2.31).
A more detailed analysis of the structure of the matrix elements (2.32)
requires further knowledge regarding the nature of the variables
$\alpha,\beta$. We will postpone the discussion of this point to later and
remark here only that orthogonality and completeness requires that

$$ \eqalign{&\sum_\alpha \int d\sigma\ \left( O^{\sigma,\alpha}_{n,p,k}\right)
   ^*O^{\sigma,\alpha}_{n',p',k'}=\delta^4(p-p')\delta^4(k-k')\delta_{nn'}
   \cr &\sum_{n=1,2} \int d^4p\int d^4k \left( O^{\sigma,\alpha}_{n,p,k}
   \right)^*O^{\sigma',\alpha'}_{n,p,k}=\delta(\sigma-\sigma')
   \delta_{\alpha,\alpha'} \cr} \eqno(2.33) $$

$$ \eqalign{&\sum_\beta \int d\sigma\ \left( O^{\sigma,\beta}_{m,p}\right)
   ^*O^{\sigma,\beta}_{m',p'}=\delta^4(p-p')\delta_{mm'}\cr
   &\sum_{m=1,2} \int d^4p\left( O^{\sigma,\beta}_{m,p}
   \right)^*O^{\sigma',\beta'}_{m,p}=\delta(\sigma-\sigma')
   \delta_{\beta,\beta'} \cr} \eqno(2.34) $$
To complete the transformation to the {\it outgoing } spectral representation
we have to calculate, according to eq. (1.22), the following quantities

$$ \langle V_m(p)\vert \Omega_+\vert \sigma,\beta\rangle_0 \qquad \langle
   N_n(p),\theta_n(k)\vert\Omega_+ \vert\sigma,\beta\rangle_0 $$
and

$$ \langle V_m(p)\vert \Omega_+\vert \sigma,\alpha\rangle_0 \qquad \langle
   N_n(p),\Theta_n(k)\vert\Omega_+ \vert\sigma,\alpha\rangle_0 $$
>From the second of equations (2.31), the discussion following eq. (2.9) and
the results of Appendix A, it is clear that the first two transformation
matrix elements are identically zero (since $\Omega_+\vert V(p)\rangle=0$). We
obtain expressions for the second pair with the help of (2.24), (2.25) and
(2.32). For the first matrix element in the second pair above we have

$$ \langle V_m(p) \vert \Omega_+ \vert \sigma,\alpha \rangle_0=$$
$$ \sum_{n=1,2}
   \int d^4p_1 \int d^4p_2 \langle V_m(p) \vert \Omega_+ \vert N_n(p_1),
   \theta_n(p_2) \rangle\langle N_n(p_1),\theta_n(p_2)\vert \sigma,\alpha
   \rangle_0=$$
$$ =\sum_{n=1,2} \int d^4p_1\int d^4p_2 \sum_{k=1,2} W^{-1}_{mk}(\omega_n
   -i\epsilon,p_1+p_2)f_{kn}(p_2)\delta^4(p-p_1-p_2)
   O^{\sigma,\alpha}_{n,p_1,p_2}=$$
$$ =\sum_{k=1,2} W^{-1}_{mk}(\sigma-i\epsilon,p) \sum_{n=1,2} \int d^4p_2\
   f_{kn}(p_2)O^{\sigma,\alpha}_{n,p-p_2,p_2}=
   \sum_{k=1,2}W^{-1}_{mk}(\sigma-i\epsilon,p)F^\alpha_k(\sigma,p)
   \eqno(2.35)$$
where we have used (2.30) and the defininition

$$ F^\alpha_k(\sigma,p) \equiv \sum_{n=1,2}\int d^4p'\ f_{kn}(p')
  O^{\sigma,\alpha}_{n,p-p',p'} \eqno(2.36)$$
For the second matrix element we get in similar way

$$ \langle N_m(p_1),\theta_m(p_2) \vert \Omega_+ \vert \sigma,
   \alpha \rangle_0=$$
$$ =\sum_{n=1,2} \int d^4p'_1\int d^4p'_2 \langle N_m(p_1), \theta_m
   (p_2) \vert \Omega_+ \vert N_n(p'_1),\theta_n(p'_2)\rangle
   \langle N_n(p'_1),\theta_n(p'_2)\vert \sigma,\alpha \rangle_0=$$
$$ =O^{\sigma,\alpha}_{m,p_1,\tilde p_2}+
   \sum_{k,j=1,2}{{f^*_{km}(\tilde p_2)W^{-1}_{kj}(\sigma-
   i\epsilon,p_1+p_2)F^\alpha_j(\sigma,p_1+p_2)}
   \over
   \sigma-i\epsilon
   -{{{p_1}^2}\over 2M_{N_m}}-{{{p_2}^2}\over 2M_{\theta_m}}}
   \eqno(2.37) $$
Following the same steps we obtain for the matrix elements of the wave
operator $\Omega_-$ 

$$ \langle V_m(p) \vert \Omega_- \vert \sigma,\alpha \rangle_0=\sum_{k=1,2}
   W^{-1}_{mk}(\sigma+i\epsilon,p)F^\alpha_k(\sigma,p) \eqno(2.38)$$
and

$$ \langle N_m(p_1),\theta_m(p_2) \vert \Omega_- \vert \sigma,
   \alpha \rangle_0
   =O^{\sigma,\alpha}_{m,p_1,p_2}+\left(\sigma+i\epsilon
   -{{{p_1}^2}\over 2M_{N_m}}-{{{p_2}^2}\over 2M_{\theta_m}}
   \right)^{-1}$$
$$ \times \sum_{k,j=1,2} f^*_{km}(p_2)W^{-1}_{kj}(\sigma+
   i\epsilon,p_1+p_2)F^\alpha_j(\sigma,p_1+p_2)
   \eqno(2.39) $$
This completes the calculation of the Lax-Phillips wave operators providing
the transformation to the {\it incoming} and {\it outgoing} (spectral)
representations. Given these transformations it is possible in principle to
construct the subspaces $D_\pm$ according to the method described in the
introduction. We can now calculate the Lax-Phillips $S$-matrix mapping
the incoming representation into the outgoing representation. If this
$S$-matrix satisfies the conditions (a),(b),(c) given in the introduction then
there exist incoming and outgoing subspaces $D_\pm$ orthogonal to each other
and the Lax-Phillips structure is complete.
\par From eq. (1.26) we see that the Lax-Phillips $S$-matrix is given by
${}_0\langle \sigma',\alpha'\vert S\vert \sigma,\alpha\rangle_0$. We can
calculate explicitly the Lax-Phillips $S$-matrix for the model presented here
with the help of the following useful expression (valid in the sector of the
Fock space in which we are working)

$$ {}_0\langle \sigma',\alpha' \vert S \vert \sigma, \alpha \rangle_0 =
   \sum_{m=1,2} \int d^4p\ {}_0\langle \sigma',\alpha' \vert \Omega_+^\dagger
   \vert V_m(p)\rangle\langle V_m(p) \vert \Omega_- \vert \sigma,\alpha
   \rangle_0$$
$$ +\sum_{m=1,2} \int d^4\tilde p_1 \int d^4\tilde p_2\  {}_0\langle \sigma'
   ,\alpha' \vert \Omega_+^\dagger \vert N_m(\tilde p_1),\theta_m(\tilde p_2)
   \rangle \langle N_m(\tilde p_1),\theta_m(\tilde p_2) \vert \Omega_-
   \vert \sigma,\alpha \rangle_0 \eqno(2.40)$$
Using the expressions obtained for the wave operators Eqs. (2.35), (2.37),
(2.38), (2.39) and the definition (2.36) we get

$$ {}_0\langle \sigma',\alpha' \vert S \vert \sigma, \alpha \rangle_0=
   \sum_{m=1,2} \int d^4p\ \sum_{k,k'=1,2} {W_{mk}^{-1}}^*(\sigma'+i
   \epsilon,p) {F_k^{\alpha'}}^*(\sigma',p)
   W_{mk'}^{-1}(\sigma+i\epsilon,p)
   {F_{k'}^\alpha}(\sigma,p)$$
$$ +\delta(\sigma'-\sigma)\delta_{\alpha\alpha'}+
   {1\over \sigma-\sigma'+i\epsilon}\int d^4p_1
   \sum_{k',j'} {F^{\alpha'}_{k'}}^*(\sigma',p_1)
   W_{k'j'}^{-1}(\sigma+i\epsilon,p_1)F_{j'}^\alpha
   (\sigma,p_1)$$
$$ +{1\over \sigma'-\sigma+i\epsilon} \int d^4p_1  
   \sum_{k,j} F^\alpha_k(\sigma,p_1)
   {W_{kj}^{-1}}^*(\sigma'+i\epsilon,p_1)
   {F_j^{\alpha'}}^*(\sigma',p_1)$$
$$ +\sum_{m=1,2} \int d^4p_1 \int d^4p_2
   \bigg[ \sum_{k,j,k',j'=1,2}
   {{f_{km}(p_2){W_{kj}^{-1}}^*(\sigma'+i\epsilon,p_1
   +p_2){F_j^{\alpha'}}^*(\sigma',p_1+p_2)}\over
   \sigma'+i\epsilon - {{{p_1}^2}\over 2M_{N_m}}-{{{p_2}^2}
   \over 2M_{\theta_m}}}$$
$$ \times{{f^*_{k'm}(p_2)W_{k'j'}^{-1}(\sigma+i\epsilon,
   p_1+p_2)F_{j'}^\alpha(\sigma,p_1+p_2)}\over
   \sigma+i\epsilon - {{{p_1}^2}\over 2M_{N_m}}-{{{p_2}^2}
   \over 2M_{\theta_m}}}\bigg] \eqno(2.41)$$
The last term in Eq. (2.41) can be put ito a simpler form by the following
manipulation

$$ \sum_{m=1,2} \int d^4p_1 \int d^4p_2
   \bigg[ \sum_{k,j,k',j'=1,2}
   {{f_{km}(p_2){W_{kj}^{-1}}^*(\sigma'+i\epsilon,p_1
   +p_2){F_j^{\alpha'}}^*(\sigma',p_1+p_2)}\over
   \sigma'+i\epsilon - {{{p_1}^2}\over 2M_{N_m}}-{{{p_2}^2}
   \over 2M_{\theta_m}}}$$
$$ \times{{f^*_{k'm}(p_2)W_{k'j'}^{-1}(\sigma+i\epsilon,
   p_1+p_2)F_{j'}^\alpha(\sigma,p_1+p_2)}\over
   \sigma+i\epsilon - {{{p_1}^2}\over 2M_{N_m}}-{{{p_2}^2}
   \over 2M_{\theta_m}}}\bigg]=$$
$$ =\int d^4p_1 \sum_{k,j,k',j'=1,2} {1\over \sigma-\sigma'}
   \left[ \delta_{kk'}(\sigma'-\sigma)-W_{kk'}(\sigma'+i\epsilon,p_1)
   +W_{kk'}(\sigma+i\epsilon,p_1) \right]$$
$$ \times
   {W_{kj}^{-1}}^*(\sigma'+i\epsilon,p_1)
   {F_j^{\alpha'}}^*(\sigma',p_1)
   W_{k'j'}^{-1}(\sigma+i\epsilon,p_1)
   F_{j'}^\alpha(\sigma,p_1)=$$
$$ =-\int d^4p_1 \sum_{k,j,j'=1,2}
   {W_{kj}^{-1}}^*(\sigma'+i\epsilon,p_1)
   {F_j^{\alpha'}}^*(\sigma',p_1)
   W_{kj'}^{-1}(\sigma+i\epsilon,p_1)
   F_{j'}^\alpha(\sigma,p_1)$$
$$ +P{1\over \sigma-\sigma'}\int d^4p_1 \sum_{j,j'=1,2}
   F_{j'}^\alpha(\sigma,p_1)
   {W_{j'j}^{-1}}^*(\sigma'+i\epsilon,p_1)
   {F_j^{\alpha'}}^*(\sigma',p_1)$$
$$ -P{1\over \sigma-\sigma'}\int d^4p_1 \sum_{j,j'=1,2}
   {F_j^{\alpha'}}^*(\sigma',p_1)
   W_{jj'}^{-1}(\sigma+i\epsilon,p_1)
   F_{j'}^\alpha(\sigma,p_1) \eqno(2.42)$$
where $P$ stands for the principle part and we have performed a partial
fraction decomposition at the second step in (2.41) and used the definition
Eq. (2.21) of $W_{ik}(z,p)$. Combining (2.42) and (2.41) we find for the
Lax-Phillips $S$-matrix
\footnote{*}{In (2.42) we use a partial fraction decomposition of the denominators of the
form $(\sigma+i\epsilon_1-A)^{-1}\times (\sigma'+i\epsilon_2-A)^{-1}=
(\sigma-\sigma'+i(\epsilon_2-\epsilon_1))^{-1}\times
((\sigma'+i\epsilon_1-A)^{-1}-(\sigma+i\epsilon_2-A)^{-1})=
(P(\sigma-\sigma')^{-1}
\pm i\pi\delta(\sigma-\sigma'))\times ((\sigma'+i\epsilon_1-A)^{-1}
-(\sigma+i\epsilon_2-A)^{-1})=P(\sigma-\sigma')^{-1}\times
 ((\sigma'+i\epsilon_1-A)^{-1}-(\sigma+i\epsilon_2-A)^{-1})$}

$$ {}_0\langle \sigma',\alpha' \vert S \vert \sigma, \alpha \rangle_0=$$
$$ =\delta(\sigma-\sigma')\bigg[ \delta_{\alpha\alpha'}
   -2\pi i \int d^4p \sum_{k,j} F^\alpha_k(\sigma,p)
   {W_{kj}^{-1}}^*(\sigma'+i\epsilon,p)
   {F_j^{\alpha'}}^*(\sigma',p_1)\bigg]$$
$$ =\delta(\sigma-\sigma')\bigg[ \delta_{\alpha\alpha'}
   -2\pi i \int d^4p \sum_{k,j}{F_j^{\alpha'}}^*
   (\sigma,p)W_{jk}^{-1}(\sigma+i\epsilon,p)
   F^\alpha_k(\sigma,p)\bigg] \eqno(2.43)$$
We observe that in eq. (2.43) the quantity ${F_j^\alpha}^*(\sigma,p)$ can be
considered, for each fixed value of $\tilde p'_1$, as a vector-valued function
on the independent variable $\sigma$, taking its values in an auxiliary
Hilbert space defined by the variables $\alpha$. We write it as
 (see equations (2.29) and (2.36))

$$ {F_j^\alpha}^*(\sigma,p) \equiv ( \vert n_j \rangle_{\sigma,p})^\alpha
   \eqno(2.44) $$
where (for a fixed value of $p$ $(\vert n_j\rangle_{\sigma,p})^\alpha$ is the
$\alpha$ component of the vector valued function
$\vert n_j\rangle_{\sigma,p}$. With this notation we have (we supress the
auxiliary Hilbert space variables $\alpha$)

$$ S(\sigma)=1-2\pi i \int d^4p \sum_{k,j}
   \vert n_j \rangle_{\sigma,p}W_{jk}^{-1}(\sigma+i\epsilon,
   p){}_{\sigma,p}\langle n_k \vert \eqno(2.45)$$
\par Further simplification of the expression given here for the $S$-matrix
can be achieved by identifying the auxiliary Hilbert space variables
$\alpha$. This results in an observation of the direct integral structure
of the $S$-matrix on the center of momentum $P$ and the definition of the
reduced $S$-matrix $S_P(\sigma)$ for each value of $P$. Another important
result is the fact that the requirement that the Lax-Phillips $S$-matrix is
an inner function implies that an analysis of its action involves a
consideration of only a two dimensional subspace of the auxiliary Hilbert
space. These simplifications in the structure of the Lax-Phillips $S$-matrix
is the subject of next section.
\bigskip
\par{\bf 3. The auxiliary Hilbert space and characterization of the
Lax-Phillips $S$-matrix}
\smallskip
\par The auxiliary Hilbert space of the Lax-Phillips representation of the
relativistic Lee-Friedrichs model aquires a complete characterization when
an exact specification of the variables $\alpha$ in the transformation
matrix $O_{n,p.k}^{\sigma,\alpha}$ of equation (2.29) is given. To achieve
this goal we proceed in two steps. The first one is to define a new set of
independent variables $\{n,p,k\}\rightarrow \{ n,P,p_{rel}\}$ by the
following linear combination of $p$ and $k$

$$ a. \qquad P=p+k \qquad b.\qquad p_{rel}= { {M_{\theta_n}p-M_{N_n}k}\over
   M_{\theta_n}+M_{N_n} }\eqno(3.1) $$
\par These momentum space variables correspond to the following configuration
space variables

$$ a. \qquad X_{c.m}={{M_{N_n}x_1+M_{\theta_n}x_2}\over M_{N_n}+M_{\theta_n}}
   \qquad b. \qquad x_{rel}=x_1-x_2 $$
\par From eq. (2.32) we know that

$$ O^{\sigma,\alpha}_{n,p,k}=\delta(\sigma-\omega_{N_n}(p)
   -\omega_{\Theta_n}(k))\tilde O^{\sigma,\alpha}_{n,p,k}$$

\par This implies that

$$ \sigma={{p^2}\over 2M_{N_n}}+{{k^2}\over 2M_{\theta_n}}={{P^2}\over 2M_n}
   +{{p_{rel}^2}\over 2\mu_n} \eqno(3.2)$$
where $M_n=M_{N_n}+M_{\theta_n}$ and $\mu=M_{N_n}M_{\theta_n}/(M_{N_n}+
M_{\theta_n})$. We take $\sigma$ and $P$ to be independent variables. In this
case $p_{rel}^2$ is a dependent variable with a value given by

$$p_{rel}^2=2\mu_n(\sigma-{{P^2}\over 2M_n})$$
\par To complete the set of independent quantum numbers we have to find a
complete set of commuting operators that commute with $p_{rel}^2$ and $P$.
Since $p_{rel}^2$ is a Casimir of the Poincare group on the relative
coordinates, we may take for the set of commuting operators on the
{\it relative motion},  the second
Casimir of the Lorentz group and $L^2$,$L_3$. We denote by $\gamma$ the
full set of quantum numbers corresponding to the latter three operators. We
then have $\{\sigma,\alpha\}\equiv \{ \sigma,n,P,\gamma\}$. It follows from
eq. (2.32) and (3.1a) that

$$ O^{\sigma,\alpha}_{n,p,k}\equiv O^{\sigma,P,\gamma,i}_{n,p,k}
   =\delta(\sigma-p^2/2M_{N_n}-k^2/2M_{\theta_n})
   \delta_{ni}\delta^4(P-p-k)
   \hat O^{n,p^2_{rel},\gamma}_{n,p_{rel}}
   \vert_{{p^2_{rel}=2\mu_n(\sigma-P^2/2M_n)} \atop p_{rel}=M_{\theta_n}p
   -M_{N_n}k)/M_n} \eqno(3.3)$$
Inserting this into the definition of $F^\alpha_k(\sigma,p)
(\equiv F_k^{P,\gamma,i}(\sigma,p))$, eq. (2.36) we get

$$ F^{P,\gamma,i}_k(\sigma,p) \equiv $$
$$ \equiv \delta^4(P-p)
   \sum_{n=1,2}\int d^4p'\ f_{kn}(p')
   \delta_{ni}
   \delta(\sigma-{{(p-p')^2}\over 2M_{N_n}}-{{p'^2}\over 2M_{\theta_n}})
   \hat O^{n,p^2_{rel},\gamma}_{n,p_{rel}}
   \vert_{{p^2_{rel}=2\mu_n(\sigma-P^2/2M_n)} \atop p_{rel}=M_{\theta_n}
   P/M_n-p'}$$
$$ =\delta^4(P-p)
   \sum_{n=1,2}\int d^4p_{rel}\ f_{kn}(M_{\theta_n}P/M_n-p_{rel})
   \delta_{ni}\delta(\sigma-{{P^2}\over 2M_n}-{{p_{rel}^2}\over 2\mu_n})
   \hat O^{n,p^2_{rel},\gamma}_{n,p_{rel}}
   \eqno(3.4)$$
We define the following $P$-dependent vector valued function

$$ (\vert n_k\rangle_{\sigma,P})^{\gamma,i}\equiv
   \sum_{n=1,2}\int d^4p_{rel}\ f^*_{kn}(M_{\theta_n}P/M_n-p_{rel})
   \delta_{ni}\delta(\sigma-{{P^2}\over 2M_n}-{{p_{rel}^2}\over 2\mu_n})
   (\hat O^{n,p^2_{rel},\gamma}_{n,p_{rel}})^*
   \eqno(3.5)$$
so that ${F_k^{P,\gamma,i}(p,\sigma)}^*=\delta^4(P-p)
(\vert n_k\rangle_{\sigma,P}^{\gamma,i}$.
When this form of $F^\alpha_k(p,\sigma)$ is used in eq. (2.43) we get

$$ {}_0\langle \sigma',\alpha' \vert S \vert \sigma, \alpha \rangle_0=
   {}_0\langle \sigma',P',\gamma',i'\vert S \vert \sigma,P,\gamma,i\rangle_0=
   \delta(\sigma'-\sigma)\delta(P'-P)S_P^{\gamma',i',\gamma,i}(\sigma)
   \eqno(3.6)$$
where we define the reduced $S$-matrix , for a specified value of the
center of momentum 4-vector $P$, to be

$$ S_P^{\gamma',i',\gamma,i}(\sigma)=\bigg[ 1-2\pi i \sum_{k,j=1,2}
   \vert n_j \rangle_{\sigma,P}W_{jk}^{-1}(\sigma+i\epsilon,
   P){}_{\sigma,P}\langle n_k \vert \bigg]^{\gamma',i',\gamma,i}\eqno(3.7)$$
\par The form of $S_P(\sigma)$ allows for a further simplification. For each
value of $\sigma$ the two vectors $\vert n_k\rangle_{\sigma,P}\ ,k=1,2$ span
a two dimensional subspace of the auxiliary Hilbert space. These vectors are,
in general, not orthogonal. We find the orthogonal projection onto the two
dimensional subspace using these non-orthogonal vectors by finding linear
combinations, denoted ${}_{\sigma,P}\langle F_i\vert$ such that

$$ {}_{\sigma,P}\langle F_i \vert n_j \rangle_{\sigma,P}=\delta_{ij}
   \eqno(3.8)$$
Denoting the projection operator on the subspace spanned by
$\vert n_k\rangle_{\sigma,P}\ ,k=1,2$ by $P_2(\sigma,P)$ we have

$$ P_2(\sigma,P)=\sum_{i=1,2} \vert n_i\rangle_{\sigma,P}\ 
   {}_{\sigma,P}\langle F_i \vert \eqno(3.9)$$
With this projection we construct the unit operator $1_{\sigma,P}$ on the
auxiliary Hilbert space and write

$$ 1_{\sigma,P}=(1_{\sigma,P}-P_2(\sigma,P))+P_2(\sigma,P) $$
Multiplying $S_P(\sigma)$ of eq. (3.7) by this unit operator from the right we
obtain (here, and in the sequal, we suppress reference to the
auxiliary Hilbert space variables $\gamma',i,\gamma,i$)

$$ S_P(\sigma)=1-P_2(\sigma,P)$$
$$ +\sum_i \bigg[\vert n_i\rangle_{\sigma,P}\ {}_{\sigma,P}\langle F_i \vert
   -2\pi i \sum_{k,j}\vert n_j \rangle_{\sigma,P}W_{jk}^{-1}(\sigma+i\epsilon,
   P){}_{\sigma,P}\langle n_k \vert n_i\rangle_{\sigma,P}\ {}_{\sigma,P}
   \langle F_i \vert\bigg]=$$
$$ =1-P_2(\sigma,P)+\sum_{i,j}\vert n_j\rangle_{\sigma,P}
   \bigg[\delta_{ji}-2\pi i \sum_k W_{jk}^{-1}(\sigma+i\epsilon,
   P){}_{\sigma,P}\langle n_k \vert n_i\rangle_{\sigma,P}\bigg]
   {}_{\sigma,P}\langle F_i \vert \eqno(3.10)$$
We now write the Kronecker delta $\delta_{ij}$ in the form
$ \delta_{ji}=\sum_k W_{jk}^{-1}(\sigma+i\epsilon,P)
W_{ki}(\sigma+i\epsilon,P)$ and get

$$ S_P(\sigma)=1-P_2(\sigma,P)+\sum_{i,j,k}\vert n_j\rangle_{\sigma,P}
   W_{jk}^{-1}(\sigma+i\epsilon,P) \bigg[
   W_{ki}(\sigma+i\epsilon,P)-2\pi i\ {}_{\sigma,p}\langle n_k
   \vert n_i\rangle_{\sigma,P}\bigg]
   {}_{\sigma,P}\langle F_i \vert \eqno(3.11)$$
In order to proceed at this point it is necessary to evaluate explicitly
the expression ${}_{\sigma,P}\langle n_k \vert n_i\rangle_{\sigma,P}$. Using
the definition Eq. (3.5) we obtain

$$ {}_{\sigma,P}\langle n_k \vert n_i\rangle_{\sigma,P}=$$
$$ =\sum_j \int d^4p_{rel}\int d^4p'_{rel}\ 
   f^*_{kj}(M_{\theta_j}P/M_j-p_{rel})f_{ij}(M_{\theta_j}P/M_j-p'_{rel})$$
$$ \times \delta(\sigma-{{P^2}\over 2M_j}-{{p_{rel}^2}\over 2\mu_j})
   \delta({{p^2_{rel}}\over 2\mu_j}-{{p^{\prime 2}_{rel}}\over 2\mu_j})
   \sum_{\gamma} (\hat O^{j,p_{rel}^2,\gamma}_{j,p_{rel}})^*
   \hat O^{j,p_{rel}^2,\gamma}_{j,p'_{rel}}$$
$$ =\sum_j \int d^4p_{rel}\ 
   \delta(\sigma-{{P^2}\over 2M_j}-{{p_{rel}^2}\over 2\mu_j})
   f^*_{kj}(M_{\theta_j}P/M_j-p_{rel})f_{ij}(M_{\theta_j}P/M_j-p_{rel})$$
$$ =\sum_j \int d^4k\ 
   \delta(\sigma-{{(P-k)^2}\over 2M_{N_j}}-{{k^2}\over 2M_{\theta_j}})
   f^*_{kj}(k)f_{ij}(k)\eqno(3.12)$$
We compare this result with the jump across the cut on the real axis of the
complex $\sigma$ plane of $W_{ki}(\sigma,P)$. With the help of the definition
eq. (2.21) we find

$$ W_{ki}(\sigma+i\epsilon,P)-W_{ki}(\sigma-i\epsilon,P)=2\pi i
   \sum_j \int d^4k\
   \delta(\sigma-{{(P-k)^2}\over 2M_{N_j}}-{{k^2}\over 2M_{\theta_j}})
   f^*_{kj}(k)f_{ij}(k)\eqno(3.13)$$
Using eq. (3.12),(3.13) we can write eq. (3.11) as

$$ S_P(\sigma)=1-P_2(\sigma,P)+\sum_{i,j,k}\vert n_j\rangle_{\sigma,P}
   W_{jk}^{-1}(\sigma+i\epsilon,P)
   W_{ki}(\sigma-i\epsilon,P){}_{\sigma,P}\langle F_i \vert \eqno(3.14)$$

\par The operator valued function $P_2(\sigma,P)$ defined in eq. (3.9) is a
projection operator for each value of $\sigma$

$$ P_2(\sigma,P)P_2(\sigma,P)=P_2(\sigma,P)$$
It is, therefore, a bounded positive operator on the real $\sigma$ axis. In
order to characterize $P_2(\sigma,P)$ we need several definitions and results
from operator theory on positive operator valued functions. We give these in
the appendix, where we prove that $P_2(\sigma,P)$ is an {\it outer function}
$^{18}$ and that it is actually independent of $\sigma$, that is

$$ P_2(\sigma,P)=P_{2,P} \eqno(3.15)$$
where $P_{2,P}$ is a projection operator on some fixed two dimensional
subspace of the auxiliary space.  This proof rests on the properties of
 $S_P(\sigma)$ as an {\it inner function}.$^{18}$ We shall assume that the
functions $f_{ij}(k)$ are such that the operator valued function
defined by Eq. $(3.14)$ has the appropriate analytic properties in
the upper half plane.
\par  In eq. (3.8) and (3.9) the vectors
$\vert n_i\rangle_{\sigma,P}$ and ${}_{\sigma,P}\langle F_i\vert$ may depend
on $\sigma$, but this dependence is such that the projection operator
$P_2(\sigma,P)$ projects on a fixed two dimensional subspace of the auxiliary
space for each and every value of $\sigma$. Eq. (3.14) can be written then in
the form

$$ S_P(\sigma)=1-P_{2,P}+\sum_{i,j,k}\vert n_j\rangle_{\sigma,P}
   W_{jk}^{-1}(\sigma+i\epsilon,P)
   W_{ki}(\sigma-i\epsilon,P){}_{\sigma,P}\langle F_i \vert \eqno(3.16)$$
and we see that the $S$-matrix $S_P(\sigma)$ acts in a non trivial way only
on a two dimensional subspace of the auxiliary space.
\par We now complete the characterization of the Lax-Phillips $S$-matrix
$S_P(\sigma)$. Eq. (3.15) implies that the projection valued function
$P_2(\sigma,P)$ projects the Hilbert space $L^2(-\infty,+\infty;H)$ on
the subspace $L^2(-\infty,+\infty;H_2)$ of vector valued functions taking
their values in some fixed two dimensional subspace $H_2$ of the auxiliary
Hilbert space. We use again the notation $P_{2,P}$ to denote the projection
$P_2(\sigma,P)$ as an operator valued function projecting on
$L^2(-\infty,+\infty;H_2)$, that is

$$ P_{2,P}: L^2(-\infty,+\infty;H) \rightarrow L^2(-\infty,+\infty;H_2)
   \eqno(3.17)$$
We denote by $P_{I-2,P}$ the operator projecting on the subspace of functions
with a range in $H \ominus H_2$. We have

$$ P_{I-2,P}: L^2(-\infty,+\infty;H) \rightarrow L^2(-\infty,+\infty;H\ominus
   H_2)\eqno(3.18)$$
It is obvious from eq. (3.17),(3.18) that

$$ L^2(-\infty,+\infty;H)=P_{2,P} L^2(-\infty,+\infty;H)
   \oplus P_{I-2,P} L^2(-\infty,+\infty;H)\eqno(3.19)$$
In particular, if $U(\tau)$ is the operator of right translation by $\tau$
units then any left translation invariant subspace $I^-_H \subset H^2_H(\Pi)$
can be written as

$$ I^-_H=P_{2,P}I^-_H\oplus P_{I-2,P}I^-_H \eqno(3.20)$$
The translation $U(\tau)$ commutes with the projections $P_{2,P}$, $P_{I-2,P}$
and, since $I^-_H$ is a left translation invariant subspace, we have
$U(\tau)I^-_H\subset I^-_H$. Denoting $I^-_{H_2}=P_{2,P}I^-_H$ we find

$$ U(\tau)I^-_{H_2}=U(\tau)P_{2,P}I^-_H=P_{2,P}U(\tau)I^-_H\subset
   P_{2,P}I^-_H= I^-_{H_2}\eqno(3.21)$$
We see that if $I^-_H$ is a left translation invariant subspace then
$I^-_{H_2}= P_{2,P}I^-_H$ is a two dimensional invariant subspace under left
translations.
\par In the Lax-Phillips theory the Lax-Phillips $S$-matrix is an inner
function that generates a left translation invariant subspace from the Hardy
class $H^2_H(\Pi)$ (this corresponds to the stability property of
${\cal D}_-$). In this case we can write

$$ I^-_H=S^{LP}H_H^2(\Pi),\eqno(3.22)$$
where $S^{LP}$ is the Lax-Phillips $S$-matrix. From eq. (3.16) we see that
in the case of the two channel relativistic Lee-model we have ($S_P(\sigma)$
is the realization of $S^{LP}$ in terms of an operator valued function)

$$ [S^{LP},P_{2,P}]=0\eqno(3.23)$$
 \par From eq. (3.22), (3.23) and the definition of $I^-_{H_2}$ we find that

$$ I^-_{H_2}=P_{2,P}I^-_H=P_{2,P}S^{LP}H^2_H(\Pi)=S^{LP}P_{2,P}H^2_H(\Pi)
   =S^{LP}H^2_{H_2}(\Pi) $$
where $H^2_{H_2}(\Pi)\equiv P_{2,P}H^2_H(\Pi)$. We can write this result in
the form

$$ I^-_{H_2}=P_{2,P}S^{LP}P_{2,P}H^2_{H_2}(\Pi)\eqno(3.24)$$
From this we see that $P_{2,P}S^{LP}P_{2,P}$, when it acts on the Hardy
space $H^2_{H_2}(\Pi)$, generates a two dimensional left translation
invariant subspace. From eq. (3.16) we get (if $A$ is an operator on a Hardy
class $H^2_H(\Pi)$ or $H^2_{H_2}(\Pi)$ then $T(A)$ is its realization in
terms of an operator valued function)

$$ T(P_{2,P}S^{LP}P_{2,P})=\delta(\sigma-\sigma')
   \sum_{i,j,k}\vert n_j\rangle_{\sigma,P}
   W_{jk}^{-1}(\sigma+i\epsilon,P)
   W_{ki}(\sigma-i\epsilon,P){}_{\sigma,P}\langle F_i \vert \eqno(3.25)$$
According to eq. (3.24) this immediately implies that the right hand side of
eq. (3.25) is an {\it inner function} acting on the Hardy space
$H^2_{H_2}(\Pi)$ consisting of vector valued functions taking their values in
some fixed two dimensional subspace of the auxiliary Hilbert space. This
observation allows for a complete characterization of the Lax-Phillips
$S$-matrix, eq. (3.16). Such an inner function can be represented as a
product of a {\it rational} inner function containing the poles and
zeros of $S_P(\sigma)$ and a factor which is an inner function with
non-vanishing determinant$^{24}$.  If the latter factor is
bounded exponentially, it corresponds to a trivial inner
factor$^{1}$ and does not change the spectrum of the
semigroup.  In the following, we consider the case of a purely rational
 $S$-matrix.
\bigskip
{\bf 4. The resonant states for a rational $S$-matrix}
\smallskip
\par In this section we shall identify the resonant states of the relativistic
two channel Lee-model in the Lax-Phillips {\it outgoing} translation
representation for the case of a rational $S$-matrix of the form

$$ S(\sigma)=1+ \left( {{{\rm Res}S(z_1)}\over \sigma - z_1} +
   {{{\rm Res}S(z_2)}\over \sigma - z_2} \right)
   \qquad {\rm Im}z_1,{\rm Im}z_2 < 0. \eqno(4.1)$$
We also have

$$ S^\dagger (\sigma)=1+ \left( {{{\rm Res}S^\dagger ({\overline z_1})}
   \over \sigma - {\overline z_1}}+{{{\rm Res}S^\dagger ({\overline z_2})}
   \over \sigma - {\overline z_2}} \right)
   \qquad {\rm Im} {\overline z_1},{\rm Im} {\overline z_2}>0
\eqno(4.2)$$
\par A rational $S$-matrix of this form implies the property, as
assumed in the remarks following Eq. $(3.15)$, that $S(\sigma)$ is an
inner factor.  There are simple conditions, which we shall discuss
elsewhere, for which the converse is true, {\it i.e.}, that an inner
function is rational.
\par In order to identify the resonant states we obtain, in the {\it outgoing}
translation representation, an expression for the generator of the
Lax-Phillips semigroup. We then find the eigenfunctions of this generator.
Lax and Phillips then assert that these are the resonant states associated with
the poles of the Lax-Phillips $S$-matrix.
\par The Lax-Phillips semigroup is defined as $Z(\tau)=P_+U(\tau)P_-,\
\tau >0$. The generator of the semigroup is given by

$$B=i \lim_{\tau\to 0^+} {{Z(\tau)-Z(0)}\over \tau} \eqno(4.3)$$
In the outgoing translation representation we have

$$ {}_{out}\langle s,\beta \vert B \vert s^\prime,\beta^\prime
   \rangle_{out}=i\lim_{\tau \to 0^+}{1\over \tau}
   \sum_{\gamma,\gamma^\prime} \int d\eta \int d\eta^\prime$$
$$ \Big[
   {}_{out}\langle s,\beta \vert P_+ \vert \eta,\gamma \rangle_{
   out}\,{}_{out}\langle \eta,\gamma \vert U(\tau)
   \vert \eta^\prime,\gamma^\prime \rangle_{out}\,
   {}_{out}\langle \eta^\prime,\gamma^\prime \vert P_-
   \vert s^\prime,\beta^\prime \rangle_{out}$$
$$ -{}_{out}\langle s,\beta \vert P_+ \vert \eta,\gamma \rangle_{out}\,
   {}_{out}\langle \eta,\gamma \vert U(0)
   \vert \eta^\prime,\gamma^\prime \rangle_{out}\,
   {}_{out}\langle \eta^\prime,\gamma^\prime \vert P_-
   \vert s^\prime,\beta^\prime \rangle_{out}
   \Big] \eqno(4.4)$$
In this representation the subspace $D_+$ is given by
$L^2(-\infty,+\infty;H)$, i.e. it defined in a simple way by its support
properties. Therefore, the operator $P_+$, the projection into the subspace
$K\oplus D_-$ is given simply by
$$ {}_{out}\langle s,\beta \vert P_+ \vert \eta,\gamma \rangle_{out}
   = \Theta (-s) \delta (s-\eta)\delta_{\gamma,\gamma^\prime}$$
Furthermore, in the outgoing translation representation the evolution is
just translation

$$ {}_{out}\langle \eta,\gamma \vert U(\tau)
   \vert \eta^\prime,\gamma^\prime \rangle_{out}=
   \delta(\eta-\tau-\eta^\prime)
   \delta_{\gamma,\gamma^\prime}$$
Then (4.4) becomes

$$ {}_{out}\langle s,\beta \vert B \vert s^\prime,\beta^\prime \rangle_{out}
   =$$
$$ =i\lim_{\tau \to 0^+}{1\over \tau}\left[ \Theta (-s)
   {}_{out}\langle s-\tau ,\beta \vert P_-
   \vert s^\prime,\beta^\prime \rangle_{out}
   -\Theta (-s) {}_{out}\langle s,\beta \vert P_-
   \vert s^\prime,\beta^\prime \rangle_{out}\right]\eqno(4.5)$$

\par We use the fact that the subspace $D_-$ is given in the incoming
translation representation in terms of its support properties. This allows us
to write

$$ P_- = \sum_\gamma \int d\eta \vert \eta,\gamma \rangle_{in} \Theta(\eta)
   \langle \eta,\gamma \vert_{in}
   = \sum_\gamma \int d\eta \Omega_- \vert \eta,\gamma \rangle_f \Theta(\eta)
   \langle \eta,\gamma \vert_f \Omega_-^\dagger \eqno(4.6)$$
In the outgoing translation representation we have

$$ {\scriptstyle out}\langle s,\beta \vert P_-
   \vert s^\prime,\beta^\prime \rangle_{out} =
   \sum_\gamma \int d\eta\,\,
   {\scriptstyle out}\langle s,\beta \vert
   \Omega_- \vert \eta,\gamma \rangle_f \Theta(\eta)
   \langle \eta,\gamma \vert_f \Omega_-^\dagger
   \vert s^\prime,\beta^\prime \rangle_{out} $$
$$ = \sum_\gamma \int d\eta\,\,
   {\scriptstyle f}\langle s,\beta \vert \Omega_+^\dagger
   \Omega_- \vert \eta,\gamma \rangle_f \Theta(\eta)
   \langle \eta,\gamma \vert_f \Omega_-^\dagger \Omega_+
   \vert s^\prime,\beta^\prime \rangle_f $$
$$ = \sum_\gamma \int d\eta\,\,
   {\scriptstyle f}\langle s,\beta \vert S
   \vert \eta,\gamma \rangle_f \Theta(\eta)
   \langle \eta,\gamma \vert_f S^\dagger
   \vert s^\prime,\beta^\prime \rangle_f $$
In this expression we would like to represent the scattering operator $S$ and
its adjoint $S^\dagger$ in the spectral representation. Performing the
appropriate Fourier transforms we get

$$ {\scriptstyle out}\langle s,\beta \vert P_-
   \vert s^\prime,\beta^\prime \rangle_{out} =
   \int d\sigma \int d\sigma^\prime \sum_\alpha \int d\eta
   e^{i\sigma s}S^{\beta,\alpha}(\sigma)e^{-i\eta \sigma}
   \Theta(\eta) e^{i\eta \sigma^\prime} {S^\dagger(\sigma^\prime)}^
   {\alpha,\beta}e^{-i\sigma^\prime s^\prime} $$
$$ ={{-i}\over 4\pi^2}
   \int d\sigma \int d\sigma^\prime \sum_\alpha e^{i\sigma s}
   {{S^{\beta,\alpha}(\sigma){S^\dagger(\sigma^\prime)}^{\alpha,\beta^\prime}}
   \over \sigma - (\sigma^\prime +i\epsilon)}
   e^{-i\sigma^\prime s^\prime} \eqno(4.7)$$
The operator valued function $S(\sigma)$ is analytic in the upper half of the
complex $\sigma$ plane. Its adjoint $S^\dagger(\sigma)$ is analytic in the
lower half plane. We assume that $S(\sigma)$ has the form (4.1) and 
has two poles in the upper half plane, located at $z_1$ and $z_2$. The poles
of $S^\dagger(\sigma)$ are thus at $\overline z_1$ and $\overline z_2$. The
form of $S(\sigma)$ and of $S^\dagger(\sigma)$ allows the integrals in (5.7)
to performed by contour integration, according to the various possible signs
of $s$ and $s'$. The result is (through the rest of this section we suppress
in our notation the auxiliary Hilbert space variables)

$$ {\scriptstyle out}\langle s \vert P_-
   \vert s^\prime\rangle_{out}=
   \Theta(s)\delta(s-s^\prime)+{1\over 2\pi} \Theta(-s)$$
$$ \times \Bigl[e^{i z_1 s}{\rm Res}\ S(z_1) \int d\sigma^\prime {{
   S^\dagger(\sigma^\prime)}\over z_1-(\sigma^\prime+i\epsilon)}
   e^{-i \sigma^\prime s^\prime} 
   +e^{i z_2 s}{\rm Res}\ S(z_2)\int d\sigma^\prime {{
   S^\dagger(\sigma^\prime)}\over z_2-(\sigma^\prime+i\epsilon)}
   e^{-i \sigma^\prime s^\prime} \Bigr]=$$
$$ \Theta(s)\delta(s-s^\prime)-i\Theta(-s)\Theta(s^\prime)
   \Bigl[e^{i z_1 s}{\rm Res}\ S(z_1) S^\dagger(z_1) 
   e^{-i z_1 s^\prime} $$
$$ +e^{i z_2 s}{\rm Res}\ S(z_2) S^\dagger(z_2)
   e^{-i z_2 s^\prime} \Bigr] $$
$$ +i \Theta(-s) \Theta(-s^\prime)
   \Bigl[ e^{i z_1 s}{\rm Res}\ S(z_1) \big(
   {{ {\rm Res}\ S^\dagger({\overline z_1})}
   \over z_1-{\overline z_1} }
   e^{-i {\overline z_1} s^\prime} 
   +{{ {\rm Res}\ S^\dagger({\overline z_2})}
   \over z_1-{\overline z_2} }
   e^{-i {\overline z_2} s^\prime} \big) $$
$$ +e^{i z_2 s}{\rm Res}\ S(z_2) \big(
      {{ {\rm Res}\ S^\dagger({\overline z_1})}
   \over z_2-{\overline z_1} }
   e^{-i {\overline z_1} s^\prime}
   +{{ {\rm Res}\ S^\dagger({\overline z_2})}
   \over z_2-{\overline z_2} }
   e^{-i {\overline z_2} s^\prime} \big) \Bigl]\eqno(4.8)$$
\par The Lax-Phillips $S$-matrix is analytic in the upper half-plane. Its
analytic continuation into the lower half-plane
is given by $S(\sigma)\equiv \left(S^\dagger(\overline\sigma)\right)^{-1},\
{\rm Im} \sigma<0$. Similarily, $S^\dagger(\sigma)$ is analytic in the lower
half-plane and its analytic continuation to the upper half-plane is given by
$S^\dagger(\sigma) \equiv (S(\sigma))^{-1},\ {\rm Im}\ \sigma >0$. At any
point in the complex plane we have

$$S(\sigma)S^\dagger(\sigma)=S^\dagger(\sigma)S(\sigma)=1\eqno(4.9)$$
This relation is obtained by analytic continuation and does not imply
unitarity of the $S$-matrix off the real axis. From (4.2) and (4.9) we
have

$$ S(\sigma)
   \left[ 1+ \left( {{{\rm Res}S^\dagger ({\overline z_1})}
   \over \sigma - {\overline z_1}}+{{{\rm Res}S^\dagger ({\overline z_2})}
   \over \sigma - {\overline z_2}} \right) \right] = 1 $$
In the limit as $\sigma$ goes to $\overline z_1$ or to $\overline z_2$ we
then get

$$ \eqalign{
   S(\sigma){\rm Res}S^\dagger({\overline z_1})&
   \simeq A_1(\sigma - {\overline z_1})
   \qquad \sigma \to {\overline z_1} \cr
   S(\sigma){\rm Res}S^\dagger({\overline z_2})&
   \simeq A_2(\sigma - {\overline z_2})
   \qquad \sigma \to {\overline z_2} \cr
   {\rm Res}S^\dagger({\overline z_1})S(\sigma)&
   \simeq \hat A_1(\sigma - {\overline z_1})
   \qquad \sigma \to {\overline z_1} \cr
   {\rm Res}S^\dagger({\overline z_2})S(\sigma)&
   \simeq \hat A_2(\sigma - {\overline z_2})
   \qquad \sigma \to {\overline z_2} \cr }\eqno(4.10)$$
for some fixed (i.e., independent of $\sigma$) operators $A_1,\ A_2,\
\hat A_1,\ \hat A_2$. From (4.1) and (4.9) we have

$$ S^\dagger(\sigma)
   \left[ 1- \left( {{{\rm Res}S(z_1)}\over \sigma - z_1} +
   {{{\rm Res}S(z_2)}\over \sigma - z_2} \right) \right]=1 $$
and in the limit as $\sigma$ approaches $z_1$ or $z_2$ we get

$$ \eqalign{
   S^\dagger(\sigma){\rm Res}S(z_1)&
   \simeq -B_1(\sigma - z_1) \qquad \sigma \to z_1 \cr
   S^\dagger(\sigma){\rm Res}S(z_2)&
   \simeq -B_2(\sigma - z_2) \qquad \sigma \to z_2 \cr
   {\rm Res}S(z_1)S^\dagger(\sigma)&
   \simeq -\hat B_1(\sigma - z_1) \qquad \sigma \to z_1 \cr
   {\rm Res}S(z_2)S^\dagger(\sigma)&
   \simeq -\hat B_2(\sigma - z_2) \qquad \sigma \to z_2 \cr
   }\eqno(4.11)$$
With the help of equations (4.11) we find that the second term of eq. (4.8)
vanishes and the representation of $P_-$ in the outgoing translation
representation to be

$$ {\scriptstyle out}\langle s,\beta \vert P_-
   \vert s^\prime,\beta^\prime \rangle_{out}=
   \Theta(s)\delta(s-s^\prime)-{1\over2\pi} \Theta(-s)$$
$$ \times \Bigl[e^{i z_1 s}{\rm Res}\ S(z_1) \int d\sigma^\prime {{
   S^\dagger(\sigma^\prime)}\over z_1-(\sigma^\prime+i\epsilon)}
   e^{-i \sigma^\prime s^\prime} 
   +e^{i z_2 s}{\rm Res}\ S(z_2)\int d\sigma^\prime {{
   S^\dagger(\sigma^\prime)}\over z_2-(\sigma^\prime+i\epsilon)}
   e^{-i \sigma^\prime s^\prime} \Bigr] $$
$$ =\Theta(s)\delta(s-s^\prime) $$
$$ +i\Theta(-s) \Theta(-s^\prime)
   \Bigl[ e^{i z_1 s}{\rm Res}\ S(z_1) \big(
   {{ {\rm Res}\ S^\dagger({\overline z_1})}
   \over z_1-{\overline z_1} }
   e^{-i {\overline z_1} s^\prime} 
   +{{ {\rm Res}\ S^\dagger({\overline z_2})}
   \over z_1-{\overline z_2} }
   e^{-i {\overline z_2} s^\prime} \big) $$
$$ +e^{i z_2 s}{\rm Res}\ S(z_2) \big(
      {{ {\rm Res}\ S^\dagger({\overline z_1})}
   \over z_2-{\overline z_1} }
   e^{-i {\overline z_1} s^\prime}
   +{{ {\rm Res}\ S^\dagger({\overline z_2})}
   \over z_2-{\overline z_2} }
   e^{-i {\overline z_2} s^\prime} \big) \Bigr] \eqno(4.12)$$
Inserting (4.12) in (4.5) we get for the generator of the semigroup

$$ {}_{out}\langle s,\beta \vert B \vert s^\prime,\beta^\prime \rangle_{out}
   =$$
$$ =i\lim_{\tau \to 0^+}{1\over \tau}\left[ \Theta (-s)
   {}_{out}\langle s-\tau ,\beta \vert P_-
   \vert s^\prime,\beta^\prime \rangle_{out}
   -\Theta (-s) {}_{out}\langle s,\beta \vert P_-
   \vert s^\prime,\beta^\prime \rangle_{out}\right]= $$
$$ =i\lim_{\tau \to 0^+}{1\over \tau}\Big\{ \Theta (-s)
   \Big(\Theta(s-\tau)\delta(s-\tau-s')$$
$$ +i \Theta(-s+\tau) \Theta(-s^\prime)
   \Bigl[e^{i z_1 (s-\tau)}{\rm Res}\ S(z_1)
   K_1(s')+ e^{i z_2 (s-\tau)}{\rm Res}\ S(z_2)K_2(s')\Bigr]\Big)$$
$$ -\Theta (-s)
   \Big(\Theta(s)\delta(s-s')
   +i \Theta(-s) \Theta(-s^\prime)
   \Bigl[e^{i z_1 s}{\rm Res}\ S(z_1)
   K_1(s') + e^{i z_2 s}{\rm Res}\ S(z_2)K_2(s')
   \Bigr]\Big)\Big\}$$
$$ =i \Theta(-s) \Theta(-s^\prime)$$
$$ \times \lim_{\tau \to 0^+}{1\over \tau}
   \Big[(e^{i z_1 (s-\tau)}-e^{i z_1 s})
   {\rm Res}\ S(z_1)K_1(s')+(e^{i z_2 (s-\tau)}-e^{i z_2 s})
   {\rm Res}\ S(z_2)K_2(s')\Big]$$
$$ =\Theta(-s) \Theta(-s^\prime)\Big[z_1\, e^{i z_1 s}
   {\rm Res}\ S(z_1)K_1(s')+ z_2\, e^{i z_2 s}
   {\rm Res}\ S(z_2)K_2(s')\Big]\eqno(4.13)$$
where we have denoted

$$ \eqalign{K_1(s')&={{ {\rm Res}\ S^\dagger({\overline z_1})}
   \over z_1-{\overline z_1} }
   e^{-i {\overline z_1} s^\prime}
   +{{ {\rm Res}\ S^\dagger({\overline z_2})}
   \over z_1-{\overline z_2} }
   e^{-i {\overline z_2} s^\prime}\cr 
   K_2(s')&={{ {\rm Res}\ S^\dagger({\overline z_1})}
   \over z_2-{\overline z_1} }
   e^{-i {\overline z_1} s^\prime}
   +{{ {\rm Res}\ S^\dagger({\overline z_2})}
   \over z_2-{\overline z_2} }
   e^{-i {\overline z_2} s^\prime}\cr}\eqno(4.14)$$
\par We show that, in the outgoing translation representation, the
eigenvectors of the generator $B$ of the Lax-Phillips semigroup are

$$\psi_1(s)=\Theta(-s){\rm Res}\ S(z_1)e^{i z_1 s},\qquad
\psi_2(s)=\Theta(-s){\rm Res}\ S(z_2)e^{i z_2 s}\eqno(4.15)$$
in the sense that a vector in $\overline H$ given by $\psi_\beta(s)=
\Theta(-s)e^{iz_1s}({\rm Res}\, S(z_1))^{\beta\beta'}\phi^{\beta'}$ or by
$\psi_\beta(s)=\Theta(-s)e^{iz_2s}({\rm Res}\, S(z_2))^{\beta\beta'}
\phi^{\beta'}$ where $\phi\in H$, is an eigenvector of the generator of the
semigroup. This is achieved by demostrating that these vectors satisfy the
eigenvalue equation

$$ \int ds^\prime {\scriptstyle out}\langle s,\beta \vert
   B \vert s^\prime,\beta^\prime \rangle_{out} \psi_{1,2}^{\beta^\prime}
   (s^\prime)= z_{1,2}\psi_{1,2}^\beta(s^\prime)\eqno(4.16)$$
We verify eq. (4.16) for $\psi_1(s)$. Inserting (4.13) into (4.16) and
performing the integration we find for the second term, containing the factor
$z_2$,

$$ {\rm Res}\ S(z_2)\int_{-\infty}^0  \big(
   {{ {\rm Res}\ S^\dagger({\overline z_1})}
   \over z_2-{\overline z_1} }
   e^{-i {\overline z_1} s^\prime} 
   +{{ {\rm Res}\ S^\dagger({\overline z_2})}
   \over z_2-{\overline z_2} }
   e^{-i {\overline z_2} s^\prime} \big)e^{i z_1 s^\prime}
   {\rm Res}\ S(z_1)=$$
$$  {i\over z_2-z_1}{\rm Res}\ S(z_2) \big(
   {{ {\rm Res}\ S^\dagger({\overline z_1})}
   \over z_2-{\overline z_1} }
   +{{ {\rm Res}\ S^\dagger({\overline z_1})}
   \over {\overline z_1}-z_1 }
   +{{ {\rm Res}\ S^\dagger({\overline z_2})}
   \over z_2-{\overline z_2} }
   +{{ {\rm Res}\ S^\dagger({\overline z_2})}
   \over {\overline z_2}-z_1 }
   \big) {\rm Res}\ S(z_1)=$$
$${\rm Res}\ S(z_2)\left(S^\dagger(z_2)-S^\dagger(z_1)\right)
  {\rm Res}\ S(z_1)=0\eqno(4.17)$$
and for the first term containing the factor $z_1$

$$ {\rm Res}\ S(z_1)\int_{-\infty}^0  \big(
   {{ {\rm Res}\ S^\dagger({\overline z_1})}
   \over z_1-{\overline z_1} }
   e^{-i {\overline z_1} s^\prime} 
   +{{ {\rm Res}\ S^\dagger({\overline z_2})}
   \over z_1-{\overline z_2} }
   e^{-i {\overline z_2} s^\prime} \big)e^{i z_1 s^\prime}
   {\rm Res}\ S(z_1)=$$
$$ -i{\rm Res}\ S(z_1)\left({ { {\rm Res}\ S^\dagger({\overline z_1})}
   \over (z_1-{\overline z_1})^2}+{ {{\rm Res}\ S^\dagger({\overline z_2})}
   \over (z_1-{\overline z_2})^2} \right) {\rm Res}\ S(z_1)=$$
$$ i{\rm Res}\ S(z_1) {{dS^\dagger(\sigma)}\over d\sigma}
   \Big\vert_{\sigma =z_1} {\rm Res}\ S(z_1)\eqno(4.18)$$
In order to simplify this last expression we need two identities, the first
of which is obtained by exploiting the unitarity of $S(\sigma)$
for real $\sigma$. Taking the derivative $d/d\sigma (S^\dagger(\sigma)
S(\sigma))$ we can write

$$ {{dS^\dagger(\sigma)}\over d\sigma}=-S^\dagger(\sigma)
   {{dS(\sigma)}\over d\sigma}S^\dagger(\sigma)\eqno(4.19)$$
The second identity is obtained with the help of equation
(5.1)

$$ {\rm Res}\ S(z_1)=S(\sigma)S^\dagger(\sigma){\rm Res}\ S(z_1)=
   \left( 1-{{{\rm Res}\ S(z_1)}\over \sigma -z_1}-{{{\rm Res}\ S(z_2)}
   \over \sigma-z_2} \right)S^\dagger(\sigma){\rm Res}\ S(z_1)$$
From this identity and eq. (4.10) we get, for small values of
$\vert \sigma-z_1\vert$

$$ {\rm Res}\ S(z_1)S^\dagger(\sigma){\rm Res}\ S(z_1) \simeq
   -{\rm Res}\ S(z_1) (\sigma - z_1)\eqno(4.20)$$
When eq. (4.19) and (4.20) are used in (4.18) we get

$$ i{\rm Res}\ S(z_1) {{dS^\dagger(\sigma)}\over d\sigma}
   \Big\vert_{\sigma =z_1} {\rm Res}\ S(z_1)=
   (-i)\lim_{\sigma \to z_1} {\rm Res}\ S(z_1)
   S^\dagger(\sigma){{dS(\sigma)}\over d\sigma}
   S^\dagger(\sigma){\rm Res}\ S(z_1)=$$
$$ =-i\lim_{\sigma \to z_1} {\rm Res}\ S(z_1)S^\dagger(\sigma)
   \left( {{{\rm Res}\ S(z_1)}\over (\sigma -z_1)^2}
   +{{{\rm Res}\ S(z_2)}\over (\sigma -z_2)^2}\right)
   S^\dagger(\sigma){\rm Res}\ S(z_1)=$$
$$ -i\lim_{\sigma \to z_1} {1\over (\sigma -z_1)^2}
   {\rm Res}\ S(z_1)S^\dagger(\sigma)
   {\rm Res}\ S(z_1)S^\dagger(\sigma){\rm Res}\ S(z_1)=
   -i {\rm Res}\ S(z_1)\eqno(4.21)$$
Making use of the results (4.21), (4.17) it is easy to
verify Eq. (4.16) for $\psi_1(s)$. A similar calculation shows that
$\psi_2(s)$ also satisfies Eq. (4.16).
\par A rational Lax-Phillips $S$-matrix is a rational, operator valued
inner function. Such an operator can be written as$^{25}$

$$ S(\sigma)={{\sigma-\overline z_1 P_1}\over \sigma- z_1 P_1}\,
             {{\sigma-\overline z_2 P_2}\over \sigma- z_2 P_2}
             \eqno(4.22)$$
where $P_1=\vert n_1\rangle\langle n_1\vert$ and $P_2=\vert n_2\rangle
\langle n_2\vert$ are projectors on one dimensional subspaces of the
auxiliary Hilbert space (we take $\vert n_1\rangle$ and $\vert n_2\rangle$ to
be normalized and that, in general, $P_1P_2\not= 0$). This $S$-matrix
can be rewritten in a form corresponding to Eq. (4.1) as

$$ S(\sigma)=1+{1\over \sigma-z_1}\big[ (z_1-\overline z_1)P_1
   {{z_1-\overline z_2P_2}\over z_1-z_2P_2}\big]
   +{1\over \sigma-z_2}\big[ {{z_2-\overline z_1P_1}\over
   z_2-z_1P_1}(z_2-\overline z_2)P_2\big]\eqno(4.23)$$
From Eq. (4.23) we we identify the two residues

$$ \eqalign{{\rm Res}\, S(z_1)&=(z_1-\overline z_1)P_1
   (1+{{z_2-\overline z_2}\over z_1-z_2}P_2)\cr
   {\rm Res}\, S(z_2)&=(1+{{z_1-\overline z_1}\over
   z_2-z_1}P_1)(z_2-\overline z_2)P_2\cr}\eqno(4.24)$$
Inserting in (4.24) the expressions for $P_1$ and $P_2$ in terms of $\vert
n_1 \rangle$ and $\vert n_2\rangle$ we find

$$ \eqalign{{\rm Res}\,S(z_1)&=(z_1-\overline z_1)\vert n_1\rangle
   \big( \langle n_1\vert +{{z_2-\overline z_2}\over z_1-z_2}
   \langle n_1\vert n_2\rangle\langle n_2\vert \big)\cr
   {\rm Res}\, S(z_2)&=\big( \vert n_2\rangle+{{z_1-\overline z_1}\over
   z_2-z_1}\vert n_1\rangle\langle n_1\vert n_2\rangle\big)(z_2-
   \overline z_2)\langle n_2\vert\cr}\eqno(4.25)$$
The eigenvectors of the generator $B$ of the semigroup, which we denote by
$\vert \chi_1 \rangle$ and $\vert \chi_2 \rangle$, can now be imediately
identified, in light of the remarks following Eq. (4.15), from Eq. (4.25).

$$\eqalign{\vert \chi_1\rangle&=\Theta(-s)(z_1-\overline z_1)\vert n_1
  \rangle e^{iz_1 s}\cr
  \vert \chi_2\rangle&=\Theta(-s)\big( \vert n_2\rangle+{{z_1-\overline
  z_1}\over z_2-z_1}\vert n_1\rangle\langle n_1\vert n_2\rangle\big)
  e^{iz_2 s}\cr}\eqno(4.26)$$
Once the residues of the $S$-matrix and the eigenvectors $\vert \chi_1\rangle$
,$\vert \chi_2\rangle$ are given explicitely in Eq. (4.25) and (4.26) we
can insert these expressions into Eq. (4.13) to achieve an explicit
expression for the generator $B$ of the semigroup. We find that

$$ B=z_1 \vert \chi_1\rangle\langle \tilde \chi_1 \vert
     +z_2 \vert \chi_2\rangle\langle \tilde \chi_2 \vert\eqno(4.27) $$
where $\langle \chi_i\vert \chi_j\rangle=\delta_{ij}$ and

$$ \eqalign{\langle \tilde \chi_1\vert&=a_1 \langle \chi_1\vert
   +b_1\langle\chi_2\vert\cr \langle \tilde \chi_2\vert&= a_2\langle \chi_1
   \vert+b_2\langle \chi_2\vert\cr}\eqno(4.28)$$
with the coefficients $a_1,b_1,a_2,b_2$ given by

$$ \eqalign{ a_1&=\vert \langle n_1\vert n_2\rangle\vert^2{{(z_1-
   \overline z_2)(\overline z_2-z_2)}\over z_1-z_2}+1\cr
   b_1&={{\overline z_2-z_2}\over z_1-z_2}
   \langle n_1\vert n_2\rangle\cr
   a_2&={{z_2-\overline z_2}\over
   \overline z_2-\overline z_1}\langle n_2\vert n_1\rangle\cr
   b_2&=\overline z_2-z_2\cr}\eqno(4.29)$$
\par Eq. (4.27) has the diagonalized form of the Lee-Oehme-Yang-Wu
phenomenological Hamiltonian in the subspace of the two
 resonance channel containing, in this case, the $K^0$ and ${\overline
 K}^0$ (or $K_S,\, K_L$) states. One sees from Eqs. $(3.12)$ and
 $(3.13)$ that the
 jump function containing the essential parameters of the $S$-matrix
 in this subspace contain the matrix elements $\{f_{ij}\}$ of the
 perturbation. These transition matrix elements coincide in form with
 the quantities calculated in quantum field theoretical models for the
 vertex for neutral $K$ meson decay. The theory that we have given
 here explains how the neutral $K$ meson corresponds to a {\it state}
 in the quantum mechanical Hilbert space (even though it is relatively
 short-lived) with an exact semigroup decay law, as seen to high accuracy in
 experiment$^{11}$. 
\bigskip
\par{\bf 5. Discussion and Conclusions}
\par We have shown that the quantum mechanical formulation of
 Lax-Phillips theory for the description of resonances and
 decay$^{5}$ can be generalized to a system with a finite discrete
 set of resonances.  If this set of resonances spans the unstable
 system subspace, the most general form of the $S$-matrix is that of a
 rational inner function$^{25}$, treated in detail in
 Section 4 for the two-dimensional case.
\par The eigenstates corresponding to the poles of the  $S$ matrix are
 well-defined vectors in the full Hilbert space ${\overline H}$, and
the left and right  eigenvectors are orthogonal with respect to the
scalar product of ${\overline H}$.  They span a two-dimensional
subspace of $\overline H$; the $S$-matrix acts non-trivially on a two
dimensional subspace of the auxiliary space $H$ for each value of the
foliation parameter $\sigma$ (independently of $\sigma$).  This
corresponds to an ideal form of ``resonance dominance.''
 \par The
relation between the eigenvectors of the generator of the semigroup
in the space $\overline H$ and the vectors spanning the
two-dimensional subspace of $H$ is very simple (see Eq. $(4.15)$). We
are therefore able to construct a model completely within the two
dimensional subspace, containing an effective non-Hermitian generator
of the semigroup, and a set of vectors in a two dimensional space with
scalar products taking the same value as the corresponding vectors in
the full space.  This two dimensional (in general, $N$-dimensional)
space and the generator of the semigroup acting on it coincides with
the Lee-Oehme-Yang-Wu model.  Moreover, as we have seen in the simple
Lee model which we have studied here, the matrix elements of the model
Hamiltonian are related to the perturbation formally in the same way as 
 in the framework of the Wigner-Weisskopf model.   
                  
\bigskip
{\bf Appendix A.}
\smallskip
\par We show that $\Omega_\pm \vert V_i(\tilde p) \rangle = 0$ applying the
methods used in section 2. The procedure is explicitly performed for
$\Omega_+ \vert V_i(\tilde p) \rangle = 0$. The result for $\Omega_- \vert
V_i(\tilde p) \rangle = 0$ is obtained in a similar way.
\par We start with the integral representation of the wave operator (see eq.
(2.10))
$$ \Omega_+ = 1+i\lim_{\epsilon \to 0} \int_0^{+\infty} U^{\dagger}(\tau)
   VU_0(\tau)e^{-\epsilon \tau} d\tau
   \eqno(A.1)$$
applying this operator to $\vert V_i(\tilde p) \rangle$ we get

$$ \Omega_+ \vert V_i(\tilde p) \rangle = \vert V(\tilde p) \rangle
   + i\lim_{\epsilon\to 0} \int_0^{+\infty} d\tau U^{\dagger}(\tau)V
   U_0(\tau)e^{-\epsilon\tau} b_i^{\dagger}(\tilde p) \vert 0 \rangle$$
$$ = \vert V_i(\tilde p) \rangle -i \lim_{\epsilon\to 0} \int_0^{-\infty}
   d\tau U(\tau)Ve^{i(\omega_{V_i}(\tilde p)-i\epsilon)\tau}
   b_i^{\dagger}(\tilde p) \vert 0 \rangle
   \eqno(A.2) $$

Where $\omega_{V_i}(p)=p^2/2M_{V_i}$. As in section 2, we want to evaluate
the time evolution in the integral and perform a Laplace transform. The
result of the action of the potential operator, given in eq. (2.8) to
$\vert V_i(p) \rangle$, is

$$ V\vert V_i(\tilde p) \rangle=V b_i^\dagger(\tilde p)\vert 0\rangle=
   \sum_{j=1,2}\int d^4k f_{ij}^*(k)
   a^\dagger_{N_j}(\tilde p-k)a^\dagger_{\theta_j}(k) \vert 0 \rangle
   \eqno(A.3) $$
A general form of a state in the sector of the Fock space with $Q_1=1$,$Q_2=0$
is given in eq. (2.20). From eq. (A.3) we find, at time $\tau=0$,
$$ A_j(q,0)=0 \qquad B_j(p,k,0)=f_{ij}^*(k)\delta^4(\tilde p-p-k)
   \eqno(A.4)$$
Defining the Laplace transformed coefficients $\tilde A_j(q,z)$ and
$\tilde B_j(p,k,z)$ as in eq. (2.18), we use eq. (2.19) and the fact that
in eq. (A.4) $A_j(q,0)=0$ to obtain

$$ \eqalign{\tilde A_l(p,z)(z-{{p^2}\over 2M_{V_l}})&=\sum_{j=1,2}
   \int d^4\, k f_{lj}(k)\tilde B_j(p-k,k,z)\cr
   \tilde B_l(p-k,k,z)(z-{{(p-k)^2}\over 2M_{N_l}}-{{k^2}\over
   2M_{\theta_l}})&= iB_l(p-k,k,0)+\sum_{j=1,2}
   f^*_{jl}(k)\tilde A_j(p,z)\cr}\eqno(A.5)$$
Solving for $\tilde A_l(p,z)$ we get

$$ \eqalign{\tilde A_l(p,z)&=i\sum_{i=1,2} W^{-1}_{lk}(z,p)\sum_{j=1,2}
   \int d^4k\, {{f_{ij}B_j(p-k,k,0)}\over
   z-{{(p-k)^2}\over 2M_{N_l}}-{{k^2}\over 2M_{\theta_l}}}\cr
   \tilde B_l(p-k,k,z)&=
   \left(z-{{(p-k)^2}\over 2M_{N_l}}-{{k^2}\over
   2M_{\theta_l}}\right)^{-1}\left[iB_l(p-k,k,0)+\sum_{j=1,2}
   f^*_{jl}(k)\tilde A_j(p,z)\right]\cr}\eqno(A.6)$$
Inserting the initial condition for $B_l(p-k,k,0)$ from eq. (A.4) in eq. (A.6)
we have

$$ \eqalign{ \tilde A_l(p,z)&=i\left[W^{-1}_{li}(p,z)
   \left(z-{{p^2}\over 2M_{V_i}}\right)-\delta_{li}\right]
   \delta^4(p-\tilde p)\cr
   \tilde B_l(p-k,k,z)&=\left(z-{{(p-k)^2}\over 2M_{N_l}}-{{k^2}\over
   2M_{\theta_l}}\right)^{-1}\sum_{j=1,2}if^*_{jl}(k)
   W^{-1}_{ji}(p,z)\left(z-{{p^2}\over 2M_{V_i}}\right)
   \delta^4(p-\tilde p)\cr}\eqno(A.7)$$
\par Performing the Laplace transform of eq. (2.15) implied by eq. (A.2), we
use the coefficients from eq. (A.7) and evaluate the resulting expression
at the point $z=\omega_{V_i}(\tilde p)-i\epsilon=\tilde p^2/2M_{V_i}
-i\epsilon$. This procedure gives the simple answer

$$ \lim_{\epsilon\to 0}\int_0^{-\infty} d\tau\, U(\tau)Ve^{i(\omega_{V_i}
   (\tilde p)-i\epsilon)}b^\dagger_i(\tilde p)\vert 0\rangle
   = -ib^\dagger_i(\tilde p)\vert 0\rangle=-i\vert V_i(\tilde p)\rangle
   \eqno(A.8)$$
and this implies the desired result.
\bigskip
{\bf Appendix B.}
\smallskip
\par We prove here that the value taken by the projection valued function
$P_{2,P}(\sigma)$ is actually a projection operator, for all values of
$\sigma$, on a fixed two dimensional subspace of the auxiliary Hilbert
space of the Lax-Phillips representation of the relativistic Lee-Friedrichs
model, this projection operator is denoted $P_{2,P}$, i.e., we prove that

$$ P_{2,P}(\sigma)=P_{2,P} $$
\par We start with the observation made at the begining of section 4 (see Eq.
(4.1) and the discussion following it) that the operator valued function
$P_{2,P}(\sigma)$, defined in eq. (3.14), is a projection operator for each
value of $\sigma$

$$P_{n,P}(\sigma) P_{n,P} (\sigma) = P_{n,P}(\sigma)
  \eqno(B.1)$$
It is, therefore, a bounded positive operator on the real $\sigma$ axis.
\par In order to proceed we need several definitions and results from
the theory of operator valued functions. We denote the upper half plane of
the complex $\sigma$ plane by $\Pi$. If
$b$ is some separable Hilbert space, we denote by ${\cal B}(b)$ the set of
bounded linear operators on $b$. We define the following sets of
${\cal B}(b)$ valued functions$^{18}$
\smallskip
{\bf Definition A}:

\item {(i)}   A holomorphic ${\cal B}(b)$ valued function $f(\sigma)$ on
      $\Pi$ is of bounded type on $\Pi$ if \break $log^+\vert f(\sigma)
      \vert_{{\cal B}(b)}$
      has a harmonic majorant on $\Pi$. The class of all such functions is
      denoted $N_{{\cal B}(b)}(\Pi)$.

\item {(ii)}  If $\phi$ is any strongly convex function, then by
      ${\cal H}_{\phi,
      {\cal B}(b)}(\Pi)$ we mean the class of all holomorphic ${\cal B}(b)$
      valued functions $f(\sigma)$ on $\Pi$ such that $\phi
      (log^+\vert f(z)\vert_{{\cal B}(b)})$ has a harmonic majorant on $\Pi$.

\item {(iii)} We define $N^+_{{\cal B}(b)}(\Pi) = \bigcup
      {\cal H}_{\phi,{\cal B}(b)}(\Pi)$, where the union is over all strongly
      convex functions $\phi$.

\item {(iv)}  By $H^\infty_{{\cal B}(b)}(\Pi)$ we mean the set of all bounded
      holomorphic ${\cal B}(b)$ valued functions on $\Pi$.

Here $log^+t={\rm max}(log t,0)$ for $t>0$ and $log0=-\infty$. The sets
$N_{{\cal B}(b)}$ and $N_{{\cal B}(b)}^+$ are called Nevanlinna
classes and ${\cal H}_{\phi,{\cal B}(b)}(\Pi)$ is a Hardy-Orlicz class.
\par We will need the following theorems and definitions:
\smallskip
{\bf Theorem A}: The following
$$ H^\infty_{{\cal B}(b)}(\Pi) \subseteq {\cal H}_{\phi,{\cal B}(b)}(\Pi)
   \subseteq N^+_{{\cal B}(b)}(\Pi) \subseteq N_{{\cal B}(b)(\Pi)} $$
is a valid sequence.
\smallskip
{\bf Definition B}: Let $u,v$ be nonzero scalar valued functions in $N^+(R)$
($N^+(R)$ is the boundary function for a scalar Nevanlinna class function).
A ${\cal B}(b)$-valued function $F$ on $R$ is of class ${\cal M}(u_i,v_i)$
if $uF,vF^* \in N^+_{{\cal B}(b)}(R)$.
\smallskip
{\bf Definition C}: If $A \in H^\infty_{{\cal B}(b)}(\Pi)$ then:

\item {(i)}  A is an {\it inner function} if the operator
$$ T(A) \colon f \to Af, \qquad f \in H^2_b(\Pi)$$
is a {\it partial isometry} on $H^2_b(\Pi)$;
\item {(ii)} A is an {\it outer function} if
$$ \bigcup \{Af \colon f \in H^2_b(\Pi) \}=H^2_M(\Pi)$$
for some subspace $M$ of $b$.
\smallskip
\par The main theorem which we will apply here is the following:
\smallskip
{\bf Theorem B}: Let $v$ be any nonzero scalar function in $N^+(R)$. If $F$ is
any nonnegative ${\cal B}(b)$-valued function of class ${\cal M}(v,v)$ on $R$
then
$$F=G^*G$$
on $R$, where $G$ is an outer function of class ${\cal M}(1,v)$ on $R$.
The factorization of $F$ is essentially unique.
\smallskip
\par As we have remarked above, we have assumed that the functions
$f_{ij}(k)$ of the Lee model are such that $S_P(\sigma)$ is an inner
function. Since $P_{n,P}(\sigma)$ is a bounded operator then, from definition
A(iv), the relation $(3.14)$ and Theorem A, we see that
$$ P_{n,P}(\sigma) \in N^+_{{\cal B}(H)}(\Pi). $$
where $H$ is the auxiliary Hilbert space of the Lax-Phillips representation of
the relativistic Lee-Friedrichs model, defined by the variables $\gamma$ in
eq. (3.3) (or eq. (3.5),(3.6) ). Furthermore, the projection operator
$P_{2,P}(\sigma)$ satisfies $(P_{2,P}(\sigma))^* = P_{2,P}(\sigma)$ and, from
definition B we immediately have
$$ P_{2,P}(\sigma) \in {\cal M}(1,1)$$
We can apply theorem B with the result that there is a unique decomposition of
$ P_{2,P}(\sigma)$ 
$$ P_{2,P}(\sigma) = G^*G=(P_{2,P}(\sigma))^* P_{2,P}(\sigma) =
   P_{2,P}(\sigma) P_{2,P}(\sigma) $$
and that $G=P_{2,P}(\sigma)$ is an {\it outer function}. We denote by $P$
the operator on $H^2_H(\Pi)$ for which the realization
is the operator valued function $P_{2,P}(\sigma)$. From definition C(ii)
we therefore have
$$ \left\{ \bigcup Pf \colon f \in H^2_H(\Pi) \right\}=
   H^2_M(\Pi) \eqno(B.2)$$
where $M$ is a subspace of the auxiliary Hilbert space $H$.
\par Now $P_{2,P}(\sigma)$ is a projection operator for each value of
$\sigma$. We have that the range of $P_{2,P}(\sigma)$ is a two dimensional
subspace of the auxiliary Hilbert space $H$ for each $\sigma$. We denote
$M(\sigma)={\rm Im}\, P_{2,P}(\sigma)$. Define

$$ \hat M=\sum_\sigma M(\sigma) $$
For each vector valued function $f\in H^2_H(\Pi)$ we have

$$P_{2,P}(\sigma)f(\sigma)\in \hat M. \eqno(B.3)$$
Furthermore, there is no subspace of $\hat M$ that has the property (B.3).
Since $(Pf)(\sigma)=P_{2,P}(\sigma)f(\sigma)\in \hat M$ for $f\in H^2_H(\Pi)$
we must have

$$ \left\{ \bigcup Pf \colon f \in H^2_H(\Pi) \right\}=
   H^2_{\hat M}(\Pi) \eqno(B.4)$$
We conclude that $\hat M$ must be a two dimensional subspace of the auxiliary
Hilbert space $H$. If it has a higher dimension we consider two different
values of $\sigma$, say $\sigma=\sigma_0$ and $\sigma=\sigma_1\not=
\sigma_0$, such that $P_{2,P}(\sigma_1)\not= P_{2,P}(\sigma_0)$. We then
take a vector $v_0\in P_{2,P}(\sigma_0)H\, , v_0\in(P_{2,P}(\sigma_1)H)^\perp$
,a scalar valued Hardy class function $g(\sigma)\in H^2(\Pi)$ and construct
the vector valued function $j(\sigma)=g(\sigma)v_0$ . Clearly,
$j \in H^2_{\hat M}(\Pi)$ (where we denote by $j$ the vector valued function
taking the value $j(\sigma)$ at the point $\sigma$) but
$j\not\in \{\cup Pf \colon f\in H^2_H(\Pi)\}$, since for any $f\in H^2_H(\Pi)$
we have $j(\sigma_1)=g(\sigma_1)v_0\perp (Pf)(\sigma_1)=P_{2,P}(\sigma_1)
f(\sigma_1)$. Therefore we have

$$ \left\{ Pf \colon f\in H^2_H(\Pi)\right\}\subset H^2_{\hat M}(\Pi)
   \eqno(B.4) $$
and we have a contradiction with eq. (B.3).
\par Since ${\rm Dim}\, \hat M=2$ we must have
$P_{2,P}(\sigma)=P_{2,P}(\sigma')$ for arbitrary $\sigma$ and $\sigma'$ and
we may write

$$ P_{2,P}(\sigma)=P_{2,P} $$
where $P_{2,P}$ is a projection operator on some fixed (independent of
$\sigma$) two dimensional subspace of $H$, which is the desired result.

\bigskip
\noindent
{\bf References}
\frenchspacing
\item{1.}P.D. Lax and R.S. Phillips, {\it Scattering Theory}, Academic
 Press, New York (1967).
\item{2.} C. Flesia and C. Piron, Helv. Phys. Acta {\bf 57}, 697
(1984).
\item{3.} L.P. Horwitz and C. Piron, Helv. Phys. Acta {\bf 66}, 694
(1993).
\item{4.} E. Eisenberg and L.P. Horwitz, in {\it Advances in Chemical
Physics}, vol. XCIX,  ed. I. Prigogine and S. Rice, Wiley, New York
(1997), p. 245.
\item{5.} Y. Strauss, L.P. Horwitz and E. Eisenberg, hep-th/9709036,
in press, Jour. Math. Phys.
\item{6.} C. Piron, {\it Foundations of Quantum Physics},
Benjamin/Cummings, Reading (1976).
\item{7.} V.F. Weisskopf and E.P. Wigner, Z.f. Phys. {\bf 63}, 54
(1930); {\bf 65}, 18 (1930).
\item{8.} L.P. Horwitz, J.P. Marchand and J. LaVita,
Jour. Math. Phys. {\bf 12}, 2537 (1971); D. Williams,
Comm. Math. Phys. {\bf 21}, 314 (1971).
\item{9} L.P. Horwitz and J.-P. Marchand, Helv. Phys. Acta {\bf 42}
1039 (1969). 
\item{10.} L.P. Horwitz and J.-P. Marchand, Rocky Mountain Jour. of
Math. {\bf 1}, 225 (1971).
\item{11.} B. Winstein, {\it et al,\ Results from the Neutral Kaon
Program at Fermilab's Meson Center Beamline, 1985-1997\/},
FERMILAB-Pub-97/087-E, published on behalf of the E731, E773 and E799
Collaborations, Fermi National Accelerator Laboratory, P.O. Box 500,
Batavia, Illinois 60510. 
\item{12.} T.D. Lee, R. Oehme and C.N. Yang, Phys. Rev. {\bf 106} ,
340 (1957).
\item{13.} T.T. Wu and C.N. Yang, Phys. Rev. Lett. {\bf 13}, 380
(1964).
\item{14.} L.P. Horwitz and L. Mizrachi, Nuovo Cimento {\bf 21A}, 625
(1974); E. Cohen and L.P. Horwitz, hep-th/9808030; hep-ph/9811332,
 submitted for publication.
\item{15.}  W. Baumgartel, Math. Nachr. {\bf 69}, 107 (1975);
 L.P. Horwitz and I.M. Sigal, Helv. Phys. Acta
 {\bf 51}, 685 (1978); G. Parravicini, V. Gorini
 and E.C.G. Sudarshan, J. Math. Phys. {\bf 21}, 2208
 (1980); A. Bohm, {\it Quantum Mechanics: Foundations
 and Applications\/,} Springer, Berlin (1986); A. Bohm,  M. Gadella
and
 G.B. Mainland, Am. J. Phys. {\bf 57}, 1105 (1989); T. Bailey and
 W.C. Schieve, Nuovo Cimento {\bf 47A}, 231 (1978).
\item{16.} I.P. Cornfield, S.V. Formin and Ya. G. Sinai,
{\it Ergodic Theory}, Springer, Berlin (1982).
\item{17.} Y. Four\`es and I.E. Segal, Trans. Am. Math. Soc. {\bf 78},
385 (1955).
\item{18.} M. Rosenblum and J. Rovnyak, {\it Hardy Classes and
Operator Theory}, Oxford University Press, New York (1985).
\item{19.} L.P. Horwitz, Found. of Phys. {\bf 25}, 39 (1995). See
also, D. Cocolicchio, Phys. Rev. {\bf D57}, 7251 (1998). The
nonrelativistic Lee model was defined in T.D. Lee, Phys. Rev.
 {\bf 95}, 1329 (1954); see also
K.O. Friedrichs, Comm. Pure and Appl. Math. {\bf 1}, 361 (1950).
\item{20.} E.C.G. Stueckelberg, Helv. Phys. Acta {\bf 14}, 322, 588
(1941); J. Schwinger, Phys. Rev. {\bf 82}, 664 (1951); R.P. Feynman,
        Rev. Mod. Phys. {\bf 20}, 367 (1948) and Phys. Rev. {\bf 80},
        440 (1950); L.P. Horwitz and C. Piron, Helv. Phys. Acta {\bf
        46}, 316 (1973); R. Fanchi, Phys. Rev {\bf D20},3108 (1979);
        A. Kyprianides, Phys. Rep. {\bf 155}, 1 (1986) (and references
 therein).
\item{21.} I. Antoniou, M. Gadella, I. Prigogine and P.P. Pronko,
        Jour. Math. Phys. {\bf 39}, 2995 (1998).
\item{22.} N. Shnerb and L.P. Horwitz, Phys. Rev. {\bf A48}, 4068
(1993); L.P. Horwitz and N. Shnerb, Found. of Phys. {\bf 28}, 1509 (1998).
\item{23.} For example, J.R. Taylor, {\it Scattering Theory},
 John Wiley and Sons,
 N.Y. (1972); R.J. Newton, {\it Scattering Theory of Particles
 and Waves}, McGraw Hill, N.Y. (1976).
\item{24.}B. Sz.-Nagy and C. Foia\c s, {\it Harmonic Analysis of
Operators on Hilbert Space}, North Holland, Amsterdam (1970).
\item{25.} P.D. Lax and R.S. Phillips, {\it Scattering Theory for
Automorphic Functions}, Princeton University Press, Princeton (1976).

\vfill
\eject
\end
\bye